\pdfoutput=1

\documentclass[fleqn,usenatbib,useAMS]{mnras}

\usepackage{graphicx}	
\usepackage{amsmath}	
\usepackage{multicol}        
\usepackage{bm}		
\usepackage{pdflscape}	
\usepackage{latexsym}
\usepackage[authoryear]{natbib}
\usepackage{bm,nicefrac}

\usepackage[T1]{fontenc}
\usepackage{ae,aecompl}

\usepackage{newtxtext,newtxmath}

\def\deg{\ifmmode^\circ\else$^\circ$\fi}

\def\Q{\ifmmode\mathcal{Q}\else$\mathcal{Q}$\fi}
\def\Mach{\ifmmode\mathcal{M}\else$\mathcal{M}$\fi}

\title[Multi-wavelength picture of HH216]
{Deciphering the Hidden Structures of HH 216 and Pillar IV in M16: Results from \emph{JWST} and \emph{HST}
}
\author[L.~K. Dewangan et al.]{L.~K. Dewangan$^{1}$\thanks{E-mail: lokeshd@prl.res.in}, 
O.~R. Jadhav$^{1,2}$,
A.~K. Maity$^{1,2}$,
N.~K. Bhadari$^{1}$, 
Saurabh Sharma$^{3}$,
\newauthor 
M. Padovani$^{4}$,
T. Baug$^{5}$, 
Y.~D. Mayya$^{6}$, and 
Rakesh Pandey$^{1}$\\
,
$^{1}$Physical Research Laboratory, Navrangpura, Ahmedabad - 380 009, India.\\
$^{2}$Indian Institute of Technology Gandhinagar Palaj, Gandhinagar 382355, India.\\ 
$^{3}$Aryabhatta Research Institute of Observational Sciences, Manora Peak, Nainital 263002, India.\\
$^{4}$INAF--Osservatorio Astrofisico di Arcetri, Largo E. Fermi 5, 50125 Firenze, Italy.\\
$^{5}$S. N. Bose National Centre for Basic Sciences, Sector-III, Salt Lake, Kolkata 700106, India.\\
$^{6}$Instituto Nacional de Astrof\'{i}sica, \'{O}ptica y Electr\'{o}nica, Luis Enrique Erro \# 1, Tonantzintla, Puebla, M\'{e}xico C.P. 72840.}

\begin{document}

\date{ }

\pagerange{\pageref{firstpage}--\pageref{lastpage}} \pubyear{2023}

\maketitle

\label{firstpage}

\begin{abstract}
To probe the star formation process, we present an observational investigation of the Pillar~IV and an ionized knot HH~216 in the Eagle Nebula (M16). 
Pillar~IV is known to host a Class~I protostar that drives a bipolar outflow. 
The outflow has produced the bow shock, HH~216, which is associated with the red-shifted outflow lobe.
The James Webb Space Telescope's  near- and mid-infrared images (resolution $\sim$0\rlap.{$''$}07 -- 0\rlap.{$''$}7) reveal the protostar as a single, isolated object (below 1000~AU). The outer boundary of Pillar~IV is depicted with the 3.3 $\mu$m Polycyclic aromatic hydrocarbon (PAH) emission. 
HH~216 is traced with the 4.05 $\mu$m Br$\alpha$ and the radio continuum emission, however it is undetected with 4.693 $\mu$m H$_{2}$ emission. HH~216 seems to be associated with both thermal and non-thermal radio emissions. High-resolution images reveal entangled ionized structures (below 3000~AU) of HH~216, which appear to be located toward termination shocks. New knots in 4.693 $\mu$m H$_{2}$ emission are detected, and are mainly found on Pillar~IV's northern side.
 This particular result supports the previously proposed episodic accretion in the powering source of HH~216. One part of the ionized jet (extent $\sim$0.16 pc) is discovered on the southern side of the driving source. 
 Using the $^{12}$CO($J$ = 1--0), $^{12}$CO($J$ = 3--2), and $^{13}$CO($J$ = 1--0) emission, observational signposts of Cloud-Cloud Collision (or interacting clouds) toward Pillar~IV are investigated. Overall, our results suggest that the interaction of molecular cloud components around 23 and 26 km s$^{-1}$ might
have influenced star formation activity in Pillar~IV.
\end{abstract}
%
\begin{keywords}
dust, extinction -- HII regions -- ISM: clouds -- ISM: individual object (HH~216) -- 
stars: formation -- stars: pre-main sequence
\end{keywords}
\section{Introduction}
\label{sec:intro}
High-resolution and high-sensitivity near-infrared (NIR) and mid-infrared (MIR) observations from the James 
Webb Space Telescope (\emph{JWST}) allow the study of the dust and gaseous structures 
around embedded protostars. This can enable us to gain a better understanding of physical processes taking place in star-forming sites \citep[e.g.,][]{pontoppidan22}. 
The \emph{JWST} facility has mapped the ``Pillars of Creation'' or ``elephant trunks'' in the Eagle Nebula (M16), which has been regarded as a site of active star formation. 
This argument is supported by the presence of one interesting and well recognized source, Herbig-Haro (HH) 216 (hereafter HH~216), that is the bow shock of a bipolar jet \citep[e.g.,][]{meaburn82,oliveira08,andersen04,flagey20}. There is also evidence of the episodic accretion in the powering source of HH~216 \citep[e.g.,][]{flagey20}. In this paper, we have re-examined the \emph{JWST} NIR and MIR images towards HH~216 (see Figure~\ref{fg1x}) to understand the underlying physical processes.

Eagle Nebula or M16 hosts the optical cluster NGC~6611, and is one of well-studied star-forming sites. 
Eagle Nebula is powered by the hot stars in the cluster NGC~6611, which are responsible for heating and ionizing the Nebula (i.e., M16 H\,{\sc ii} region). 
On the basis of GAIA parallax observations, \citet{kuhn19} estimated a distance of 1.74$\pm$0.13 kpc to NGC~6611, which is used in this paper \citep[see also][]{karim2023sofia}. 
The collision between two molecular cloud components has been proposed to explain the existence of O-type stars in this cluster \citep{nishimura21}. 
Using the Hubble Space Telescope (HST) data, it is investigated that M16 contains Pillars of neutral gas, which have been referred to as ``Pillars of Creation'' or ``elephant trunks'' \citep[see][for more details]{hester96}.
These Pillars \citep[labeled as I, II, III, and IV; see Figure~1 in][]{oliveira08} are associated with signs of ongoing star formation \citep[e.g.,][]{white99,indebetouw07,oliveira08,sofue20}. Using the photometry data at 1--8 $\mu$m, \citet{indebetouw07} identified candidate young stellar objects (YSOs) in the Eagle Nebula. They suggested that the Pillars might have been influenced by the feedback of massive stars, however spontaneous star formation was proposed in the dense parts of these Pillars.


The interesting object HH~216 (or M16-HH1) was identified in the H$\alpha$ radial velocity map \citep{meaburn82,andersen04,flagey20} and also in the radio 8.7 GHz continuum emission \citep[see Figure~2d in][]{white99}. Using the optical nebular emission lines (i.e., [O\,{\sc ii}]$\lambda$3726,29, H$\beta$, [O\,{\sc iii}]$\lambda$4959,5007, [N\,{\sc ii}]$\lambda$6548,84, H$\alpha$, 
and [S\,{\sc ii}]$\lambda$6717,31) around HH~216, \citet{flagey20} detected several knots, bipolar structure of the flow (extent $\sim$1.8 pc at a distance of 2.0 kpc), and 
likely powering source \citep[see also][]{andersen04,indebetouw07,oliveira08,karim2023sofia}. These results together were considered as signatures of the episodic accretion in 
the driving source of the HH~216 outﬂow that is located at the tip of the Pillar~IV \citep[see Figure~13 in][]{flagey20}. 
Apart from the embedded source, two objects (i.e., HH-N and HH-S) are also identified toward the Pillar~IV \citep[see Table~1 in][]{indebetouw07}. 
\citet{healy04} reported a water maser emission (see Table~3 in their paper) toward the embedded source that is found between the red lobe (radial velocity range: [15, 50] km s$^{-1}$) and the blue lobe (radial velocity range: [$-$50, $-$15] km s$^{-1}$; see Figure~13 in \citet{flagey20}) investigated using the H$\alpha$ radial velocity map. \citet{linsky07} did not detect X-ray emission toward the driving source, however a weak X-ray source was found at the location of HH~216. Figure~\ref{fg1x}a displays the dust continuum map at 850 $\mu$m, where the locations of Pillars~I--IV and HH~216 are marked. Figure~\ref{fg1x}b presents the {\it Spitzer} 4.5 $\mu$m image overlaid with the positions of HH-N, HH-S, bipolar structure (i.e., red and blue lobes), and Pillar~IV \citep[see][]{white99,flagey20}. 
The radio 8.71 GHz continuum emission contours are shown in Figure~\ref{fg1x}c, where the location of HH~216 is also indicated. 

This paper focuses on studying the formation process of the driving source of the bow shock HH~216 and its internal structure, which are yet to be carried out.  
Furthermore, the nature of the radio emission associated with the HH~216 outﬂow (see Figure~\ref{fg1x}a) is also examined using multi-frequency data. Molecular line data are employed to study the distribution and kinematics of the molecular gas toward Pillar~IV, which hosts the driving source of HH~216.  

Section~\ref{sec:obser} presents various observational data sets used in this paper. 
The outcomes of this paper are presented in Section~\ref{sec:data}. 
Section~\ref{sec:disc} contains a thorough discussion of our observational findings. 
Finally, Section~\ref{sec:conc} gives the conclusions of the present work.
\section{Data sets}
\label{sec:obser}
We utilized observational data sets at different wavelengths (see Table~\ref{tab1}) toward 
an area (size $\sim$3\rlap.{$'$}82 $\times$ 6\rlap.{$'$}14; central coordinates: $\alpha_{2000}$ = 18$^{h}$18$^{m}$55\rlap.$^{s}$39 and $\delta_{2000}$ = $-$13$\degr$51$'$37\rlap.{$''$}52) highlighted by the dot-dashed box in Figure~\ref{fg1x}a.  

The \emph{HST} F657N image (Proposal ID: 13926; PI: Zolt Levay) was downloaded from the MAST archive. 
We obtained the level-3 science ready \emph{JWST} NIRCam and MIRI images (Proposal ID: 2739; Proposal PI: Pontoppidan, Klaus M. ) from the MAST archive. 
The \emph{JWST} facility had observed these images at broad, medium, and narrow band filters. 
More details of \emph{JWST} performance can be found in \citet{rigby2023}. 

We examined the {\it Herschel} column density and dust temperature maps (resolution $\sim$12$''$) in this work, which were generated for the {\it EU-funded ViaLactea project} \citep{Molinari10b} using the Bayesian Point Process Mapping ({\it PPMAP}) procedure \citep{marsh15,marsh17} on the {\it Herschel} continuum images at 70--500 $\mu$m \citep{Molinari10a}. 

We also used the FUGIN $^{12}$CO($J$ = 1$-$0) and $^{13}$CO($J$ = 1$-$0) line data (plate scale = 8\rlap.{$''$}5 pixel$^{-1}$), which are calibrated in main beam temperature \citep[$T_\mathrm{mb}$, see][]{umemoto17}. According to \citet{umemoto17}, the typical RMS noise levels ($\sigma$) for $^{12}$CO and $^{13}$CO lines are $\sim$1.5~K and $\sim$0.7~K, respectively. The FUGIN line data have a velocity resolution of $\sim$1.3 km s$^{-1}$. A Gaussian function with full width at half maximum (FWHM) of 3 pixels (i.e., 25.5$''$) was used to smooth these data sets. Hence, the resultant angular resolution of these data sets is $\sim$33$''$.

The processed James Clerk Maxwell Telescope's (JCMT's) $^{12}$CO($J$ = 3--2) spectral data cube (rest frequency = 345.7959899 GHz) and the SCUBA-2 850 $\mu$m continuum map of the object ``g16'' (proposal id: M16AP088) were downloaded from the JCMT Science Archive/Canadian Astronomy Data Centre (CADC). 
The JCMT line data are calibrated in antenna temperature ($T_\mathrm{A}$). The line observations were obtained for an integration time of 1398 s, using the Heterodyne Array Receiver Programme/Auto-Correlation Spectral Imaging System \citep[HARP/ACSIS;][]{buckle09} spectral imaging system. 
The beam size, plate scale, and $\sigma$ values for the data cube are $\sim$14$''$, $\sim$7\rlap.{$''$}3, and, $\sim$1 K, respectively. 
The JCMT CO line data cube had a channel width of $\sim$0.0265 km s$^{-1}$.  
Note that the JCMT line data are employed only for an area containing the object HH~216 and the Pillar~IV.
\begin{table*}
\scriptsize
\setlength{\tabcolsep}{0.1in}
\centering
\caption{Table provides information about the observational data sets used in this work.}
\label{tab1}
\begin{tabular}{lcccr}
\hline 
  Survey/facility  &  Wavelength/      &  Resolution        &  Reference \\   
    &  Frequency/line(s)       &   ($\arcsec$)        &   \\   
\hline
\hline
NRAO VLA Archive Survey (NVAS)                 &4.89~GHz, 8.71~GHz                     & $\sim$1.55 and $\sim$9.45        &\citet{crossley07}\\
FOREST Unbiased Galactic plane Imaging survey with the Nobeyama 45-m 
telescope (FUGIN) survey   & $^{12}$CO(J = 1--0), $^{13}$CO(J = 1--0) & $\sim$20 and $\sim$21        &\citet{umemoto17}\\
James Clerk Maxwell Telescope (JCMT)     & 850 $\mu$m, CO(J = 3--2)  & $\sim$14 & PROJECT code = M16AP088; PI: Zhiyuan Ren\\
\emph{JWST} ERO MIRI F770W, F1130W, F1500W imaging facility & 7.7, 11.3, 15 $\mu$m                  & $\sim$0.44--0.70          &\citet{2015PASP..127..584R,wright2015_miri}\\ 
{\it Spitzer} Galactic Legacy Infrared Mid-Plane Survey Extraordinaire (GLIMPSE)       &3.6, 4.5, 5.8, 8.0  $\mu$m                   & $\sim$2           &\citet{benjamin03,glimpse1}\\
\emph{JWST} ERO NIRCam Short Wavelength (SW) F090W, F187N, F200W imaging facility & 0.901, 1.874, 1.99 $\mu$m                   
& $\sim$0.07          &\citet{2005SPIE.5904....1R,2012SPIE.8442E..2NB}\\ 
\emph{JWST} ERO NIRCam Long Wavelength (LW) F335M, F444W, F470N imaging facility & 3.365, 4.421, 4.707 $\mu$m                   
& $\sim$0.17          &\citet{2005SPIE.5904....1R,2012SPIE.8442E..2NB}\\ 
Hubble Space Telescope (HST) Wide Field Camera 3 (WFC3)/UVIS F657N imaging facility                & 6563 \AA + 6583 \AA                     & $\sim$0.067--0.156         &\citet{hester96}\\
\hline          
\end{tabular}
\end{table*}
\section{Results}
\label{sec:data}
\subsection{Multi-scale and multi-wavelength view of HH~216}
\label{sec:phy_env}
\begin{figure*}
\center
\includegraphics[width=\textwidth]{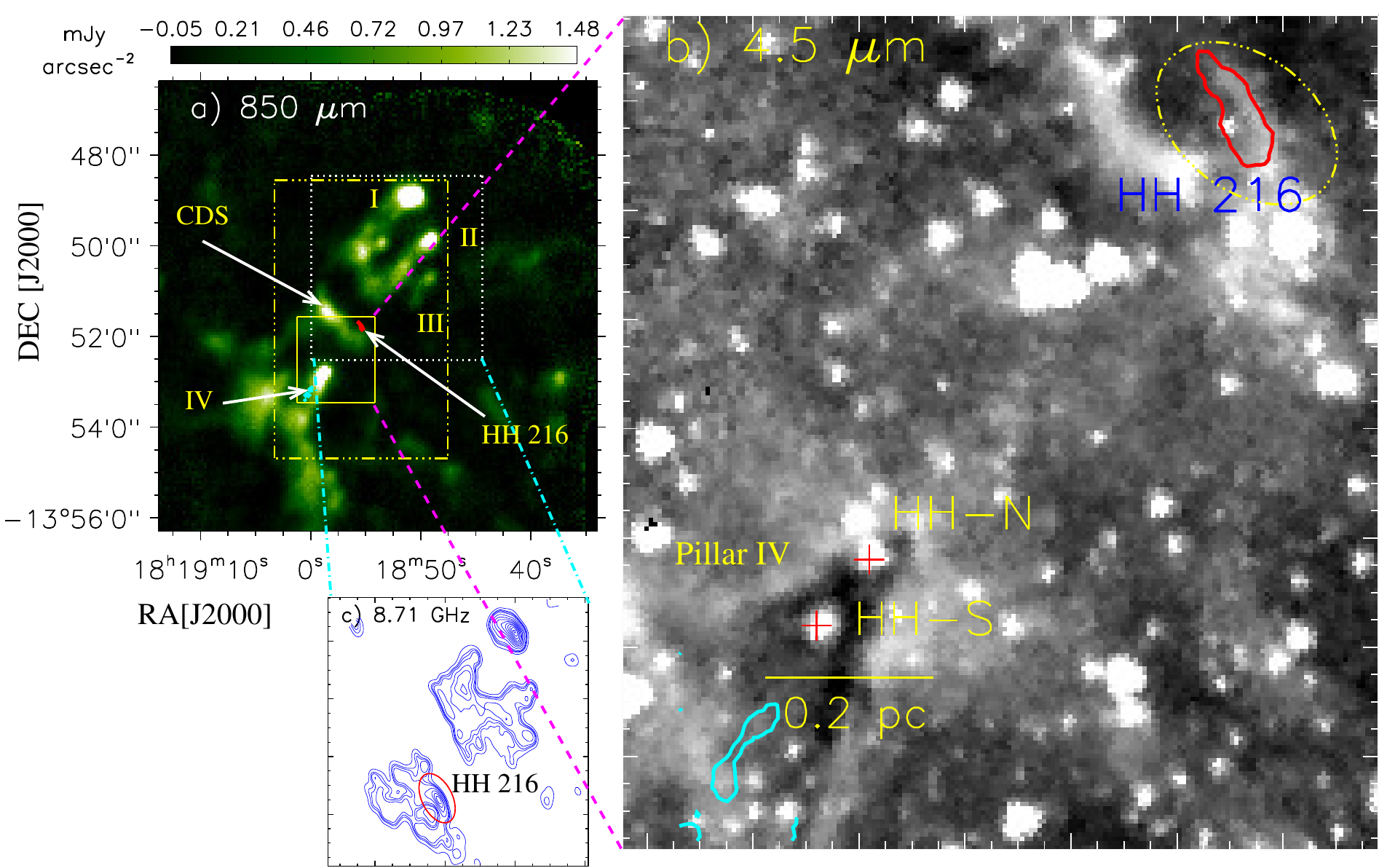}
\caption{850 $\mu$m dust emission map of the ``Pillars of Creation'' (also known as Eagle Nebula, or Messier 16) obtained using SCUBA-2 on the JCMT. 
Pillars, HH~216, and ``central dense structure'' (``CDS'') are labeled. 
The solid box highlights the area shown in Figure~\ref{fg1x}b, while the dotted box encompasses the area presented in Figure~\ref{fg1x}c.
The dot-dashed box highlights the area shown in Figures~\ref{fg2x}a and~\ref{fg2x}b. 
b) {\it Spitzer} 4.5 $\mu$m image. The outflow signature (i.e., red and cyan lobes from the H$\alpha$ radial velocity map; see curves and also \citet{flagey20}) is indicated. 
A scale bar corresponding to 0.2 pc (at a distance of 1.74 kpc) and two objects (i.e., HH-N and HH-S; see cross symbols) are presented. 
c) Radio 8.71 GHz continuum emission contours. 
The levels of the contours are 30.5 mJy beam$^{-1}$ $\times$ (0.036, 0.07, 0.09, 0.12, 0.2, 0.3, 0.4, 0.5, 0.6, 0.7, 0.8, 0.9, and 0.98). 
Pillar~IV is indicated in panels ``a'' and ``b''. In panels ``b'' and ``c'', the ellipse shows the location of HH~216.}
\label{fg1x}
\end{figure*}

As mentioned earlier, Figures~\ref{fg1x}a,~\ref{fg1x}b,and~\ref{fg1x}c highlight some previously known features/results toward M16, which are related to a bipolar outflow, an interesting object HH~216, HH-N, HH-S, and Pillars~I--IV. 
We have also labeled a dense region as ``central dense structure'' (hereafter ``CDS''), which is located between the base of Pillar~II and the top of Pillar~IV (see Figure~\ref{fg1x}a). 
Note that this paper does not focus on the Pillars~I--III and ``CDS''. 
The locations of the red lobe at V$_{\mathrm{lsr}}$ = [15, 50] km s$^{-1}$ and the blue lobe at V$_{\mathrm{lsr}}$ = [$-$50, $-$15] km s$^{-1}$ \citep[from][]{flagey20} are also marked in Figure~\ref{fg1x}b. The object HH~216 coincides with the red lobe, and is also associated with the radio continuum emission at 8.71 GHz (see the ellipse in Figure~\ref{fg1x}c).
 
The {\it Spitzer}-GLIMPSE images at 3.6--5.8 $\mu$m show a point-like source, G016.9105+00.7199 (at $\alpha_{2000}$ = 18$^{h}$18$^{m}$58\rlap.$^{s}$9, and $\delta_{2000}$ = $-$13$\degr$52$'$46\rlap.{$''$}97) located close to HH-N (Figures~\ref{fg1x}b) and its photometric magnitudes are 
m$_{3.6}$ = 12.47$\pm$0.23, m$_{4.5}$ = 11.19$\pm$0.16, and m$_{5.8}$ = 9.93$\pm$0.24 \citep[see][]{spitzer09}. 
This point-like object is the probable driving source of the outflow \citep[see also][]{andersen04,indebetouw07,oliveira08,flagey20}. Following the color conditions (i.e., [m$_{4.5}$$-$m$_{5.8}$] $\ge$ 0.7 and [m$_{3.6}$$-$m$_{4.5}$] $\ge$ 0.7) reported in \citet{hartmann05} and \citet{getman07}, this object is identified as a Class~I protostar candidate. Considering these photometric magnitudes, we also carried out the spectral energy distribution of this particular object \citep[see][for more details]{indebetouw07}, favouring that it is a candidate low-mass object ($\sim$2--3 M$_{\odot}$).

In this paper, high resolution \emph{JWST} NIRCaM and MIRI images are examined to study the sub-structures toward HH~216 and Pillar~IV. 
Previously, \citet{reiter22} examined the \emph{JWST} F470N$-$F444W image to trace the H$_{2}$ emission at 4.693 $\mu$m in NGC 3324. 
Furthermore, \citet{dewangan17a} employed the {\it Spitzer} ratio map of 4.5 $\mu$m/3.6 $\mu$m emission to study the signs of molecular outﬂows 
and the impact of massive stars on their surroundings in Sh 2-237. Nearly similar point response functions (PRFs) of the {\it Spitzer} 3.6 and 4.5 $\mu$m 
images allow them to create the {\it Spitzer} ratio map. We have also generated the F470N$-$F444W image and the ratio map of F444W and F335M using the \emph{JWST} NIR images \citep[see also][]{dewangan2023new}.

\begin{figure*}
\center
\includegraphics[width=\textwidth]{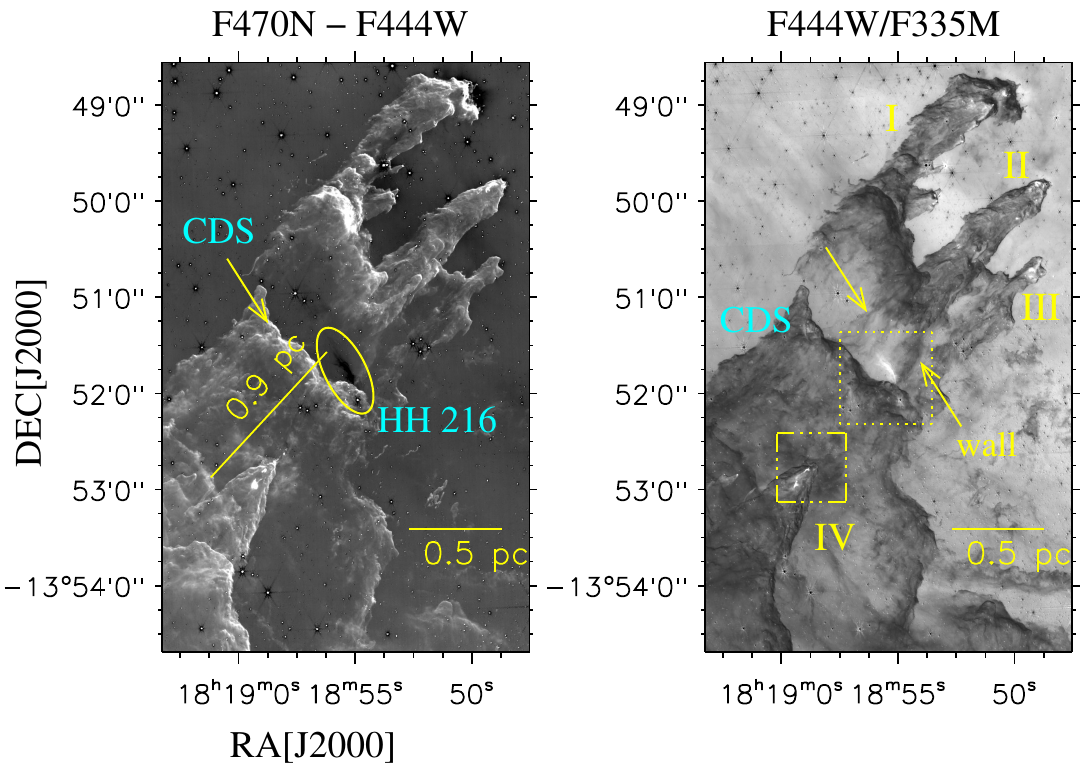}
\caption{a) \emph{JWST} F470N$-$F444W (in linear scale; see the dot-dashed box in Figure~\ref{fg1x}a). 
The ellipse shows the location of HH~216 as in Figure~\ref{fg1x}c. ``CDS'' is also indicated by an arrow. 
b) \emph{JWST} F444W/F335M (in linear scale). The dotted box and the dot-dashed box encompass the area presented in Figures~\ref{fg3x} and~\ref{fg6x}, respectively. Arrows highlight ``walls'' of the Pillar~II. 
In each panel, the \emph{JWST} image is processed through median filtering with a width of 6 pixels and smoothing by 3 $\times$ 3 pixels using the ``box-car'' 
algorithm. In all panels, a scale bar corresponding to 0.5 pc (at a distance of 1.74 kpc) is drawn.}
\label{fg2x}
\end{figure*}

Figure~\ref{fg2x}a displays the \emph{JWST} F470N$-$F444W image, revealing the regions with the H$_{2}$ emission at 4.693 $\mu$m. 
The H$_{2}$ emission is evident toward HH-N and HH-S in Pillar~IV. The object HH~216 is seen in the direction of the dark gray/black areas in Figure~\ref{fg2x}a. 
In Figure~\ref{fg2x}b, we present the \emph{JWST} ratio map (i.e., F444W/F335M) of F444W ($\lambda_{eff}$/$\Delta$$\lambda$: 4.421/1.024 $\mu$m) and F335M ($\lambda_{eff}$/$\Delta$$\lambda$: 3.365/0.347 $\mu$m) images, indicating the presence of the 3.3 $\mu$m polycyclic aromatic hydrocarbon (PAH) feature (see dark gray/black areas). 
Here one can note that the \emph{JWST} NIRCam F444W filter lacks the 3.3 $\mu$m PAH feature. The areas with the noticeable 4.05 $\mu$m Br$\alpha$ feature and/or the 4.693 $\mu$m H$_{2}$ emission seem to be identified by the bright areas in the \emph{JWST} ratio map. 
The regions with the 4.05 $\mu$m Br$\alpha$ emission in the \emph{JWST} ratio map can be easily found by looking at the locations with the 4.693 $\mu$m H$_{2}$ emission seen in the \emph{JWST} F470N$-$F444W image. 
On the basis of this approach, the Br$\alpha$ emission is traced toward the object HH~216, HH-N, and HH-S. 
The 3.3 $\mu$m PAH feature is seen toward ``CDS'' and outer boundaries/walls of Pillars I--IV. 
Additionally, the boundaries/walls of Pillar~II may also be seen extending in the southern direction, where one end of HH~216 appears to connect to one of the boundaries/walls of Pillar~II (see Figure~\ref{fg2x}b). 

\begin{figure*}
\center
\includegraphics[width=\textwidth]{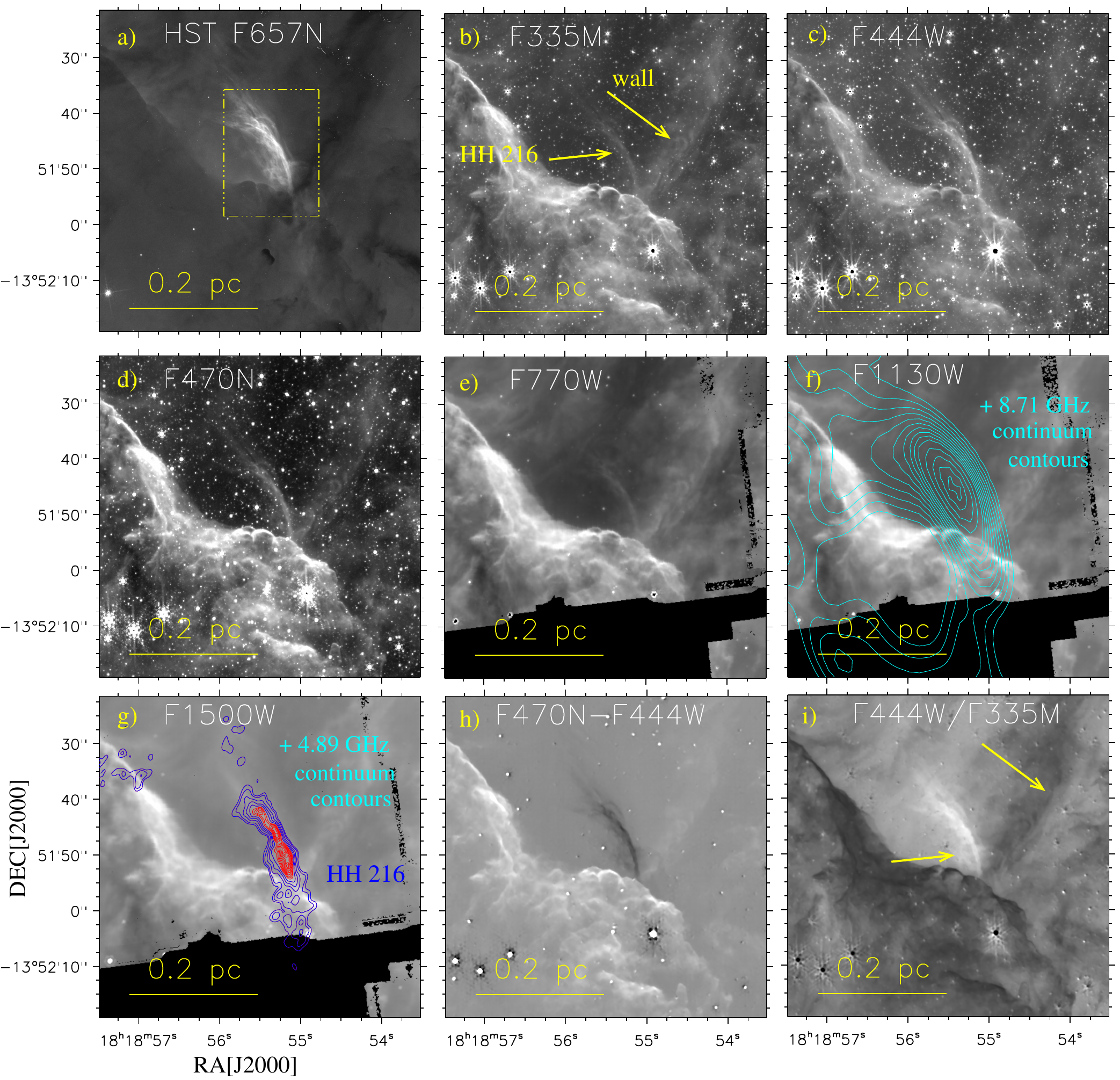}
\caption{Multi-wavelength view of the area hosting HH~216 (see the dotted box in Figure~\ref{fg2x}b). 
a) \emph{HST} F657N image. The dot-dashed box encompasses the area presented in Figures~\ref{fg5x}a--\ref{fg5x}c. 
b--g) \emph{JWST} F335M, F444W, F470N, F770W, F1130W, and F1500W images are displayed, respectively. 
h) F470N$-$F444W (in linear scale). i) F444W/F335M (in linear scale). 
In panel ``f'', the 8.71 GHz radio continuum emission contours (in cyan) are also shown and the contour levels are 17.22 mJy beam$^{-1}$ $\times$ (0.15, 0.2, 0.25, 0.3, 0.35, 0.4, 0.45, 0.5, 0.55, 0.6, 0.7, 0.8, 0.9, and 0.98). In panel ``g,'' the 4.89 GHz radio continuum emission contours (in blue and red) are also overlaid on the \emph{JWST} F1500W image. 
The blue contours are shown with the levels of 1.23 mJy beam$^{-1}$ $\times$ (0.15, 0.2, 0.3, 0.4, and 0.5), 
while the red contours are presented with the levels of 1.23 mJy beam$^{-1}$ $\times$ (0.6, 0.65, 0.7, 0.75, 0.8, 0.85, 0.9, 0.95, and 0.98). 
In each panel, a scale bar corresponding to 0.2 pc (at a distance of 1.74 kpc) is drawn.}
\label{fg3x}
\end{figure*}

Using the \emph{JWST} NIR \& MIR, \emph{HST} F658N, and NVAS radio continuum images, a small area hosting HH~216 is presented 
in Figures~\ref{fg3x}a--\ref{fg3x}g. The NVAS radio 8.71~GHz continuum emission contours (rms $\sim$1.84 mJy/beam) are overlaid on the F1130W image (see Figure~\ref{fg3x}f), while the F1500W image is overlaid with the NVAS radio~4.89~GHz continuum emission contours (rms $\sim$60.2 $\mu$Jy/beam; see Figure~\ref{fg3x}g). For a comparison purpose, the \emph{JWST} F470N$-$F444W and F444W/F335M images are also shown in Figures~\ref{fg3x}h and~\ref{fg3x}i, respectively. In Figures~\ref{fg3x}b and~\ref{fg3x}i, arrows indicate the Pillar~II's wall/boundary and HH~216. The object HH~216 is associated with the ionized emission as traced in the radio continuum maps, \emph{JWST} ratio map, and \emph{HST} F658N image. These data sets also suggest the presence of sub-structures in HH~216. 

\begin{figure*}
\center
\includegraphics[width=\textwidth]{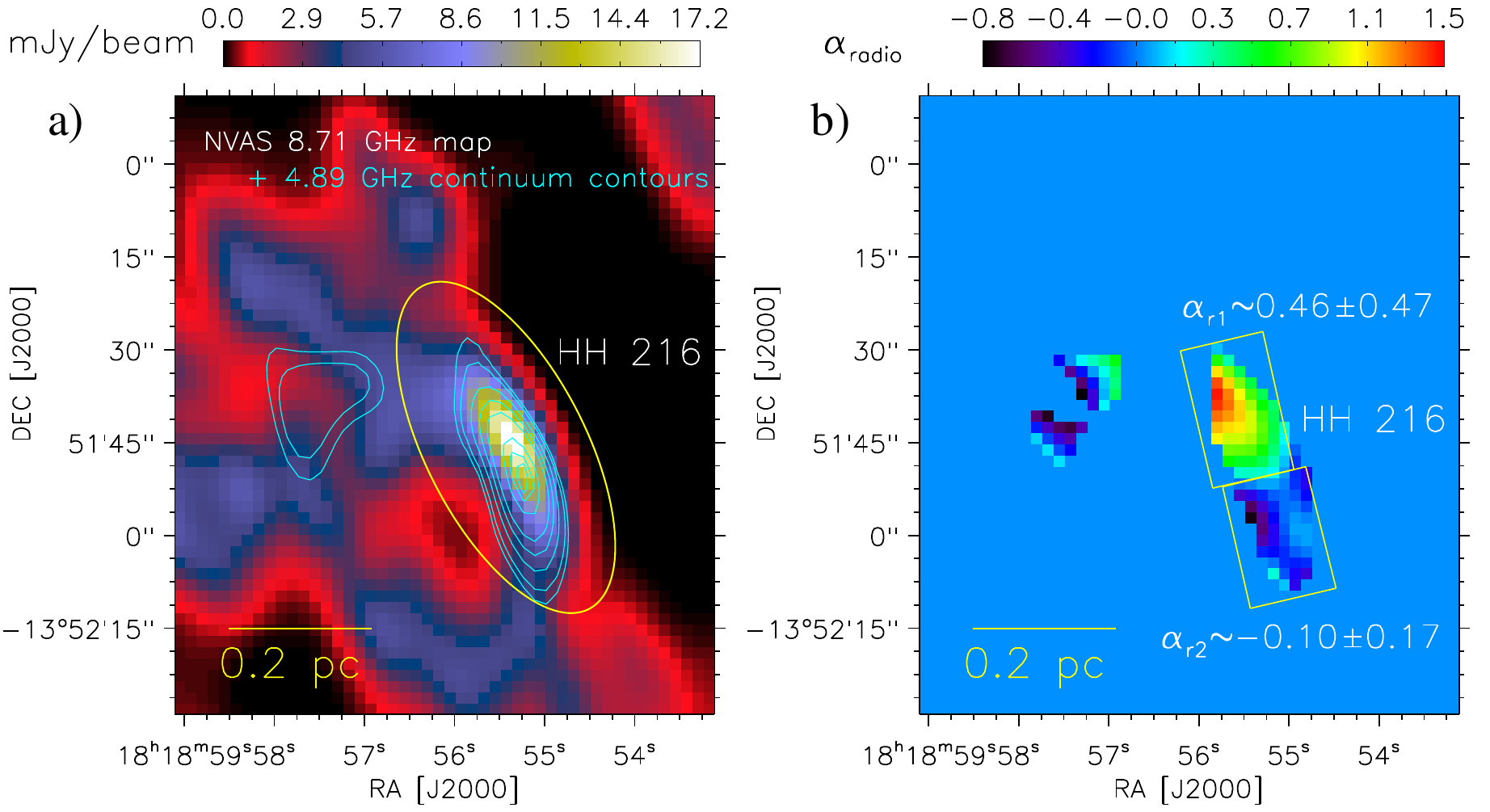}
\caption{a) Overlay of the NVAS 4.89 GHz radio continuum contours on the 8.71 GHz radio continuum map in the direction of HH~216. 
The beam size of both the radio maps is $\sim$11\rlap.{$''$}84 $\times$ 7\rlap.{$''$}07 (or $\sim$9\rlap.{$''$}45). 
The cyan contours are shown with the levels of 12.9 mJy beam$^{-1}$ $\times$ (0.25, 0.3, 0.4, 0.5, 0.6, 0.8, and 0.95). 
The ellipse indicates the location of HH~216. 
b) Radio spectral index map (resolution $\sim$11\rlap.{$''$}84 $\times$ 7\rlap.{$''$}07) of M16 produced for pixels above 
the 3$\sigma$ level in the 4.89 and 8.71 GHz bands. In the direction of HH~216, two small boxes are plotted, 
and an average value of the spectral index for each box is also indicated in the panel. 
In each panel, a scale bar corresponding to 0.2 pc (at a distance of 1.74 kpc) is drawn.}
\label{fg4x}
\end{figure*}

In order to infer the nature of radio continuum emission associated with the object HH~216, we have computed a radio spectral index ($\alpha$) map using the NVAS maps at 4.89 and 8.71 GHz. In general, a relation, F$_\nu$ $\propto$ $\nu^{\alpha}$, is known, where $\nu$ is the frequency of observation and F$_\nu$ is the corresponding observed flux density. 
A positive spectral index (i.e., $\alpha$ $>$ 0) indicates that the emission has a thermal origin, whereas a negative spectral index (i.e., $\alpha$ $<$ 0) shows the presence of non-thermal emission \citep[e.g.,][]{rybicki79,longair92}. 
In this relation, firstly, the radio map at 4.89 GHz (beam size $\sim$1\rlap.{$''$}85 $\times$ 1\rlap.{$''$}26 or $\sim$1\rlap.{$''$}55) was convolved to the beam size of the radio map at 8.71 GHz (i.e., 11\rlap.{$''$}84 $\times$ 7\rlap.{$''$}07 or $\sim$9\rlap.{$''$}45) using the CASA ``imsmooth'' task, and was also regridded to the pixel size of the radio map at 8.71 GHz ($\sim$1\rlap.{$''$}82) using the IDL function ``hastrom''. 
In Figure~\ref{fg4x}a, we have overlaid the 4.89~GHz radio continuum emission contours on the radio continuum map at 8.71~GHz. 
Using the total fluxes at 4.89 and 8.71~GHz of HH~216, we have computed an average value of the spectral index to be 0.33$\pm$0.15. 

Furthermore, we employed a pixel-wise fitting approach for the flux densities from these two radio maps, resulting in a radio spectral index map (see Figure~\ref{fg4x}b). 
In the direction of HH~216, a variation in the spectral index can be observed, ranging from $-$0.74 to 1.45. This variation represents a smooth gradient from negative to positive values as we move toward the northeast direction.
This finding allows us to select two small areas/boxes (i.e., ``r1'' and ``r2''; see Figure~\ref{fg4x}b) toward HH~216. 
The mean spectral index values are determined to be 0.46$\pm$0.47 and $-$0.10$\pm$0.17 for 
the top box (i.e., ``r1'') and the bottom box (i.e., ``r2''), respectively. This exercise seems to suggest the presence of a mix of both thermal 
and non-thermal radio emission toward HH~216. It is noted that the area ``r2'' having a negative spectral index is located toward the Pillar~II's wall/boundary. 
Further exploration of the non-thermal radio emission toward HH~216, low-frequency radio continuum observations (i.e., below 800 MHz) will be helpful.

\begin{figure*}
\center
\includegraphics[width=\textwidth]{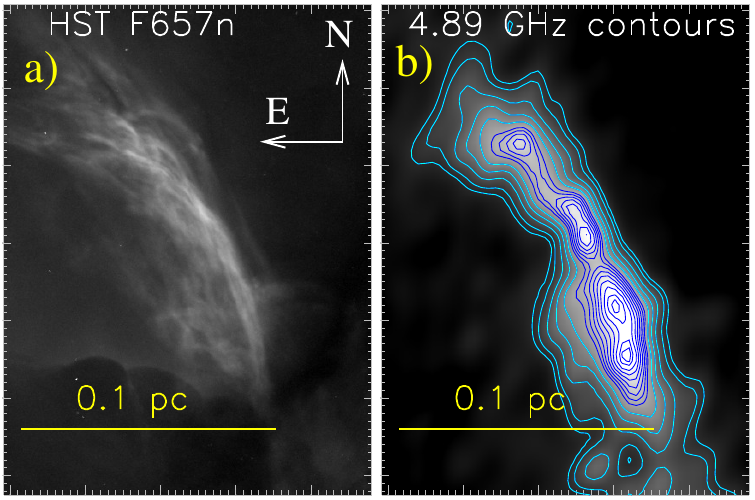}
\caption{A zoomed-in view of HH~216 using the multi-wavelength images (see the dot-dashed box in Figure~\ref{fg3x}a). 
a) \emph{HST} F657N image. b) NVAS 4.89 GHz radio continuum emission map (beam size $\sim$1\rlap.{$''$}55) and contours. 
Contours (in cyan and blue) are the same as presented in Figure~\ref{fg3x}g. 
In each panel, a scale bar corresponding to 0.1 pc (at a distance of 1.74 kpc) is plotted.}
\label{fg5x}
\end{figure*}

To study sub-structures, a zoomed-in view of the object HH~216 is presented in Figures~\ref{fg5x}a and~\ref{fg5x}b. Figure~\ref{fg5x}a displays the \emph{HST} F657N image. The NVAS 4.89 GHz continuum emission map (beam size $\sim$1\rlap.{$''$}55) and contours is presented in Figure~\ref{fg5x}b. Several ionized or radio continuum peaks are seen toward HH~216. 
From Figure~\ref{fg5x}a, we find at least two sub-structures (below 3000~AU) associated with ionized emission toward HH~216, which appear to be intertwined/entangled. 
Such configuration or entangled sub-structures may be also seen in the \emph{JWST} F470N$-$F444W image (see Figure~\ref{fg3x}h) and the \emph{JWST} F444W/F335M image (see Figure~\ref{fg3x}i). 

\begin{figure*}
\center
\includegraphics[width=\textwidth]{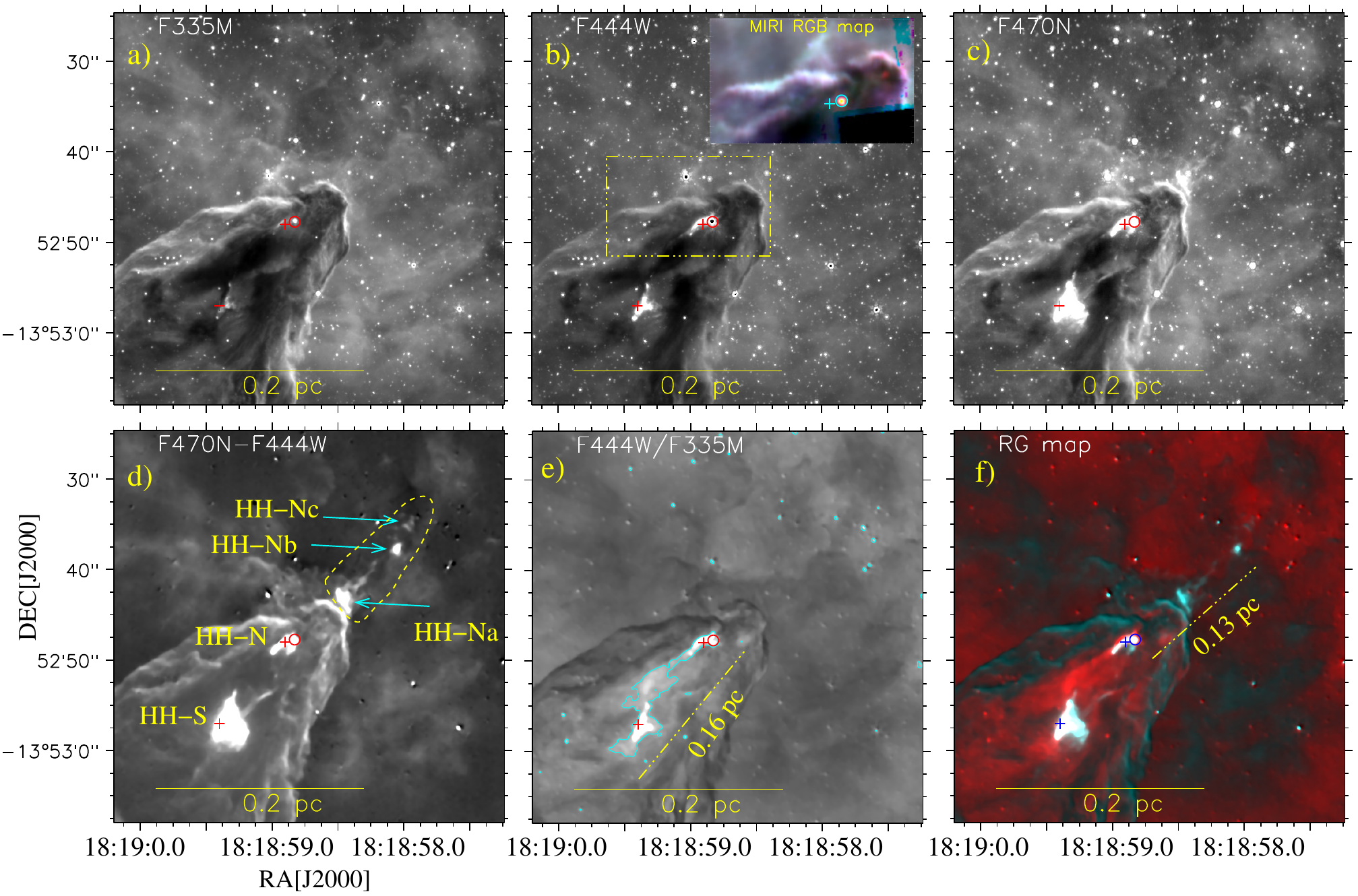}
\caption{A zoomed-in view of Pillar~IV using the multi-wavelength images (see the dot-dashed box in Figure~\ref{fg2x}b). 
a) \emph{JWST} F335M image. b) \emph{JWST} F444W image. An inset on the top right presents a zoomed-in view of the central region 
(see the dot-dashed box in Figure~\ref{fg6x}b). 
The inset is a three-color composite map made using the F1500W (in red), F1130W (in green), 
and F770W (in blue) images. c) \emph{JWST} F470N image. 
d) F470N$-$F444W image (in linear scale). Some noticeable H$_{2}$ knots are also labeled and highlighted in the panel. e) F444W/F335M image (in linear scale). 
A contour (in cyan) highlights an elongated emission feature traced in the F444W/F335M image. 
f) A two-color composite map made using F444W/F335M (in red) and F470N$-$F444W (in turquoise). 
In each panel, cross symbols are the same as shown in Figure~\ref{fg1x}b, and the circle represents the location of the driving source of the previously reported outflow. 
In all panels, a scale bar corresponding to 0.2 pc (at a distance of 1.74 kpc) is plotted.}
\label{fg6x}
\end{figure*}

As highlighted earlier, Pillar~IV hosts HH-N, HH-S, and the driving source of the bipolar outflow. 
To examine the embedded environment of Pillar~IV, using the \emph{JWST} images, a zoomed-in view is displayed in Figure~\ref{fg6x}. 
The \emph{JWST} F335M, F444W, F470N, F470N$-$F444W, and F444W/F335M images are shown in Figures~\ref{fg6x}a,~\ref{fg6x}b,~\ref{fg6x}c,~\ref{fg6x}d, and~\ref{fg6x}e, respectively. Figure~\ref{fg6x}f presents a two color composite image using the F444W/F335M (in red) and F470N$-$F444W (in turquoise) images. 
The \emph{JWST} images at $\lambda$ $\ge$ 2 $\mu$m have revealed the driving source of the bipolar outflow ($\alpha_{2000}$ = 18$^{h}$18$^{m}$58\rlap.$^{s}$8; $\delta_{2000}$ = $-$13$\degr$52$'$47\rlap.{$''$}7), which is at an offset of 1\rlap.{$''$5 from the position of the {\it Spitzer} source, G016.9105+00.7199. The locations of HH-N, HH-S, and the powering source are indicated in each panel of Figure~\ref{fg6x}. Using the \emph{JWST} MIR RGB map (F1500W (in red), F1130W (in green), and F770W (in blue) images), in Figure~\ref{fg6x}b, the inset shows a zoom in view to an area hosting the driving source. 
This source is saturated in the \emph{JWST} F444W and F770W images, and appears as a single object (below 1000 AU) in the \emph{JWST} images.
The H$_{2}$ emission at 4.693 $\mu$m is traced toward HH-N, HH-S, and the tip of the Pillar~IV (see Figures~\ref{fg6x}c and~\ref{fg6x}d). Apart from the objects HH-N and HH-S, we have also highlighted three new H$_{2}$ features in Figure~\ref{fg6x}d, which are HH-Na, HH-Nb, and HH-Nc. A separation between the driving source and the object HH-Nc is $\sim$0.13 pc (see Figure~\ref{fg6x}f)}.  
We determine that the sky-projected distance between HH-Na and HH-Nb is $\sim$13789.5~AU, but the separation between HH-Nb and HH-Nc is $\sim$6264~AU.  
The boundaries of Pillar~IV are clearly depicted in the H$_{2}$ and PAH emission (see Figures~\ref{fg6x}d and~\ref{fg6x}e). In Figures~\ref{fg6x}e and~\ref{fg6x}f, the 4.05 $\mu$m Br$\alpha$ emission is also detected toward both HH-N and HH-S, but no 4.05 $\mu$m Br$\alpha$ emission is found toward the tip of the Pillar~IV (see HH-Na, HH-Nb, and HH-Nc). 
Interestingly, an elongated feature (extent $\sim$ 0.16~pc) containing HH-N and HH-S at its opposite edges is also traced 
with the Br$\alpha$ emission (see a cyan contour in Figure~\ref{fg6x}e). This could be one part of the ionized jet, which is located on the southern side of the driving source.  
\subsection{Kinematics of molecular gas}
\label{sec:coem} 
\subsubsection{FUGIN molecular line data}
\label{Asec:coem} 
\begin{figure*}
\center
\includegraphics[width=\textwidth]{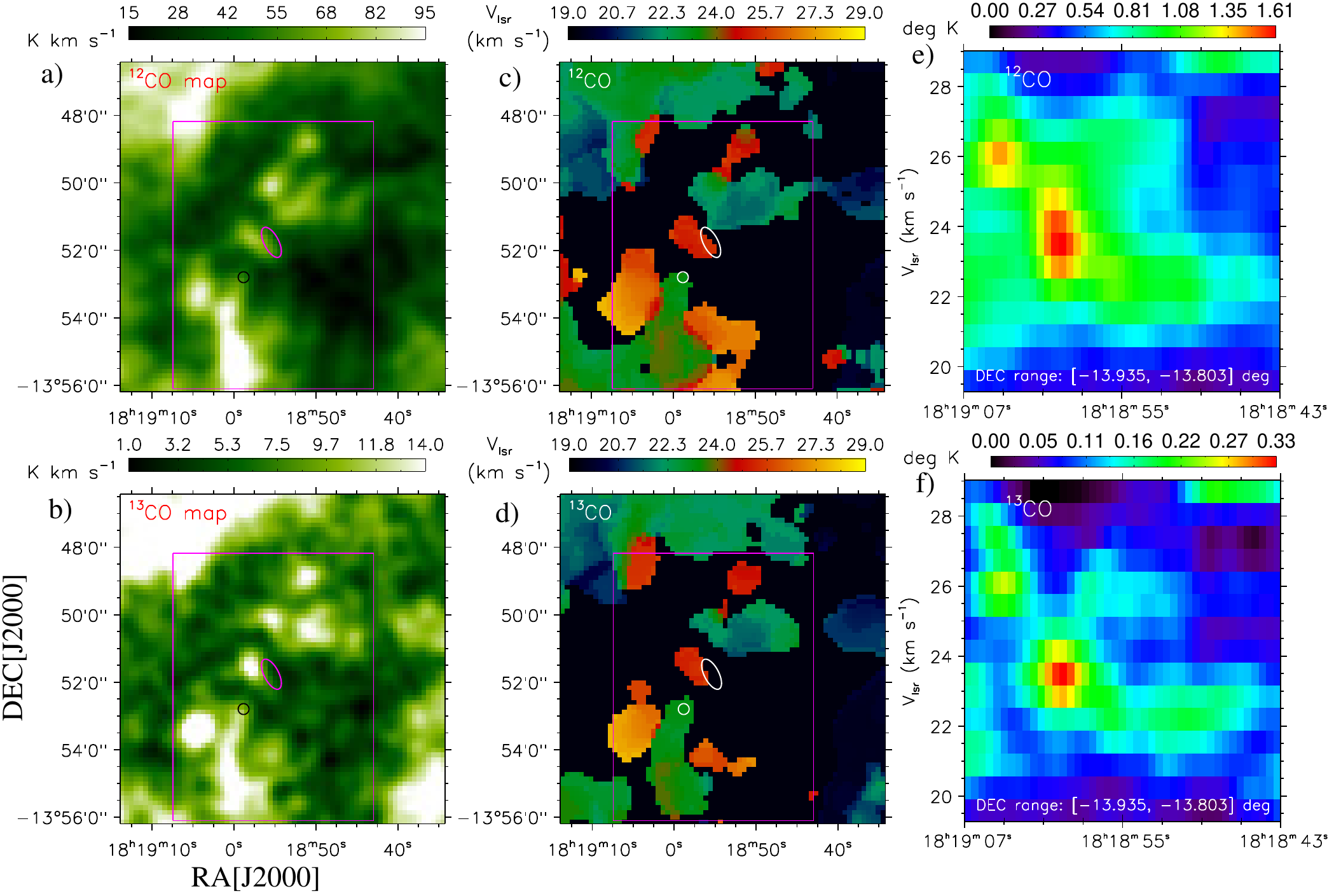}
\caption{a) FUGIN $^{12}$CO($J$ = 1--0) integrated intensity map (at [19.275, 27.725] km s$^{-1}$). 
b) FUGIN $^{13}$CO($J$ = 1--0) integrated intensity map. 
c) The line-of-sight intensity weighted velocity map of the FUGIN $^{12}$CO($J$ = 1--0) emission. 
d) FUGIN $^{13}$CO($J$ = 1--0) intensity weighted velocity map. 
e) Position (or right ascension)-velocity diagram of the FUGIN $^{12}$CO($J$ = 1--0) emission. 
The molecular emission is integrated over the declination range from $-$13.935 degrees (or $-$13$\degr$56$'$06\rlap.{$''$}0) to $-$13.803 degrees (or $-$13$\degr$48$'$10\rlap.{$''$}8) degrees (see a solid box in Figures~\ref{fg7x}a--\ref{fg7x}d). 
f) Same as Figure~\ref{fg7x}e, but for the FUGIN $^{13}$CO($J$ = 1--0) emission. 
In panels ``a--d'', the circle represents the location of the driving source 
of the previously reported outflow (see Figure~\ref{fg6x}), and the ellipse shows the location of HH~216. 
The selected area in panels ``a--d'' is the same as presented in Figure~\ref{fg1x}a. A solid box highlights an area containing major structures in M16 (see panels ``a--d'').}
\label{fg7x}
\end{figure*}

This section deals with the study of molecular line data toward Pillar~IV and its surroundings in M16. 
We employed the FUGIN $^{12}$CO($J$ = 1--0) and $^{13}$CO($J$ = 1--0) line data, and the molecular emission is studied 
in a velocity range of [19.275, 27.725] km s$^{-1}$. 
To study the distribution of molecular gas, the integrated intensity map, the line-of-sight intensity weighted velocity map, and the position-velocity diagram are produced using the molecular (i.e., $^{12}$CO($J$ = 1--0) and $^{13}$CO($J$ = 1--0)) line data.
Integrated intensity maps of $^{12}$CO($J$ = 1--0) and $^{13}$CO($J$ = 1--0) are presented in Figures~\ref{fg7x}a and~\ref{fg7x}b, respectively. Molecular gas is traced toward the previously known Pillars~I--IV and ``CDS''. Figures~\ref{fg7x}c and~\ref{fg7x}d show intensity weighted velocity maps of $^{12}$CO($J$ = 1--0) and $^{13}$CO($J$ = 1--0), respectively.
Both the intensity weighted velocity maps hint at the presence of two velocity components toward M16. 

\begin{figure*}
\center
\includegraphics[width=\textwidth]{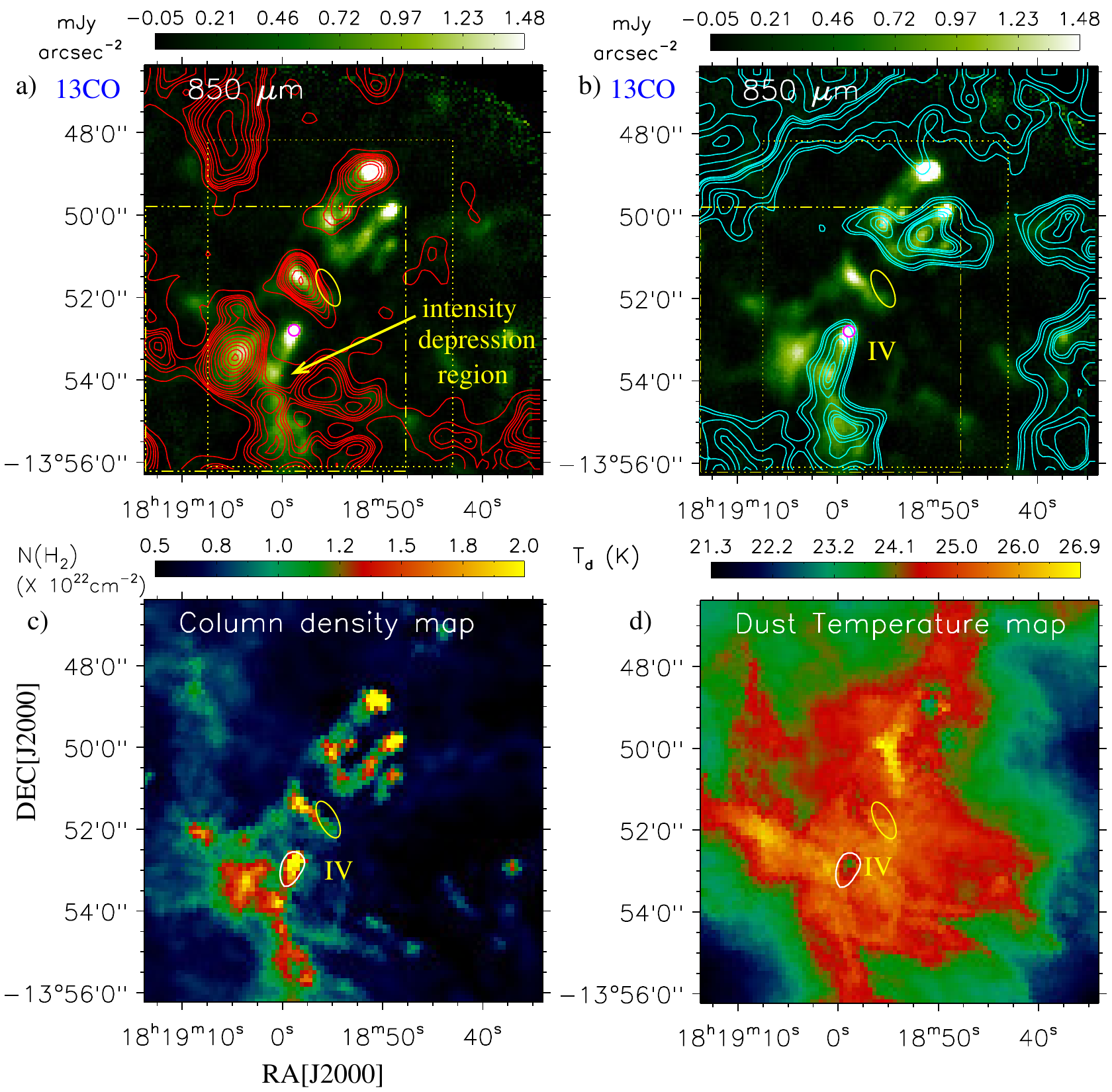}
\caption{a) Overlay of the $^{13}$CO(J= 1--0) emission contours (in red; at [24.475, 27.725] km s$^{-1}$) on the 850 $\mu$m continuum emission map. 
The contours are 18.01 K km s$^{-1}$$\times$ (0.15, 0.2, 0.22, 0.24, 0.27, 0.3, 0.35, 0.4, 0.5, 0.6, 0.7, 0.8, 0.9, and 0.98). 
b) Overlay of the $^{13}$CO(J= 1--0) emission contours (in cyan; at [19.275, 23.825] km s$^{-1}$) on the 850 $\mu$m continuum emission map. 
The contours are 17.76 K km s$^{-1}$ $\times$ (0.32, 0.35, 0.4, 0.5, 0.55, 0.58, 0.6, 0.7, 0.8, 0.9, and 0.98). 
c) {\it Herschel} column density map. d) {\it Herschel} dust temperature map. 
In panels ``a'' and ``b'', boxes highlight areas containing major structures in M16. In panels ``c'' and ``d, a contour (in white) with $N({{\rm{H}}}_{2})$ = 1.3  $\times$ 10$^{22}$ cm$^{-2}$ is also shown.}
\label{hfg8x}
\end{figure*}

Figures~\ref{fg7x}e and~\ref{fg7x}f present position (or right ascension)-velocity diagrams of $^{12}$CO($J$ = 1--0) and $^{13}$CO($J$ = 1--0), respectively, 
which reveal two velocity components around 23 and 26 km s$^{-1}$ and their connections in velocity. To produce this position-velocity diagram, we integrated the molecular emission over the declination range from $-$13.935 to $-$13.803 degrees. 
On the basis of the position-velocity diagrams, we have selected velocity ranges of two cloud components, which are [19.275, 23.825] and [24.475, 27.725] km s$^{-1}$. Using the $^{12}$CO($J$ = 1--0) and $^{13}$CO($J$ = 1--0) line data, the spatial distribution of these two cloud components and their overlapping areas is examined (not shown here). 
Using the $^{13}$CO(J= 1--0) emission, the cloud components at [24.475, 27.725] and [19.275, 23.825] km s$^{-1}$ are presented 
in Figures~\ref{hfg8x}a and~\ref{hfg8x}b, respectively. 
In Figures~\ref{hfg8x}a and~\ref{hfg8x}b, the background is the JCMT 850 $\mu$m continuum map as presented in Figure~\ref{fg1x}a. 
The Pillar~IV is associated with the blue-shifted cloud at [19.275, 23.825] km s$^{-1}$ (see Figure~\ref{hfg8x}b), which seems 
to be spatially located toward the intensity-depression region in the red-shifted cloud at [24.475, 27.725] km s$^{-1}$ (see an arrow in Figure~\ref{hfg8x}a). 
Such configuration shows a complementary distribution of the two cloud components toward the Pillar~IV, 
which can be considered as an interesting finding. Additionally, we find that the molecular gas associated with the blue-shifted cloud component is also seen toward the Pillars~II and~III (see Figure~\ref{hfg8x}b). These outcomes are also found in the $^{12}$CO(J= 1--0) emission (not shown here).

Figure~\ref{hfg8x}c displays the {\it Herschel} column density ($N({{\rm{H}}}_{2})$) map, showing the presence of high values (1--3.6 $\times$ 10$^{22}$ cm$^{-2}$) toward Pillar~IV. In the direction of Pillar~IV, a contour with $N({{\rm{H}}}_{2})$ = 1.3 $\times$ 10$^{22}$ cm$^{-2}$ is also plotted in Figure~\ref{hfg8x}c (see also Figure~\ref{hfg8x}d). 
Following the analysis presented in \citet{dewangan20g}, the {\it Herschel} column density map enables us to 
compute the total mass of Pillar~IV to be $\sim$36 $M_{\odot}$. In Figure~\ref{hfg8x}d, we present the {\it Herschel} dust temperature ($T_\mathrm{d}$) map. 
The Pillar~IV is found with $T_\mathrm{d}$ $\sim$24~K, and is surrounded by warm dust emission ($T_\mathrm{d}$ $\sim$25--27~K; see the white contour in Figure~\ref{hfg8x}d). 
\subsubsection{JCMT molecular line data}
\label{Bsec:coem} 
\begin{figure*}
\center
\includegraphics[width=\textwidth]{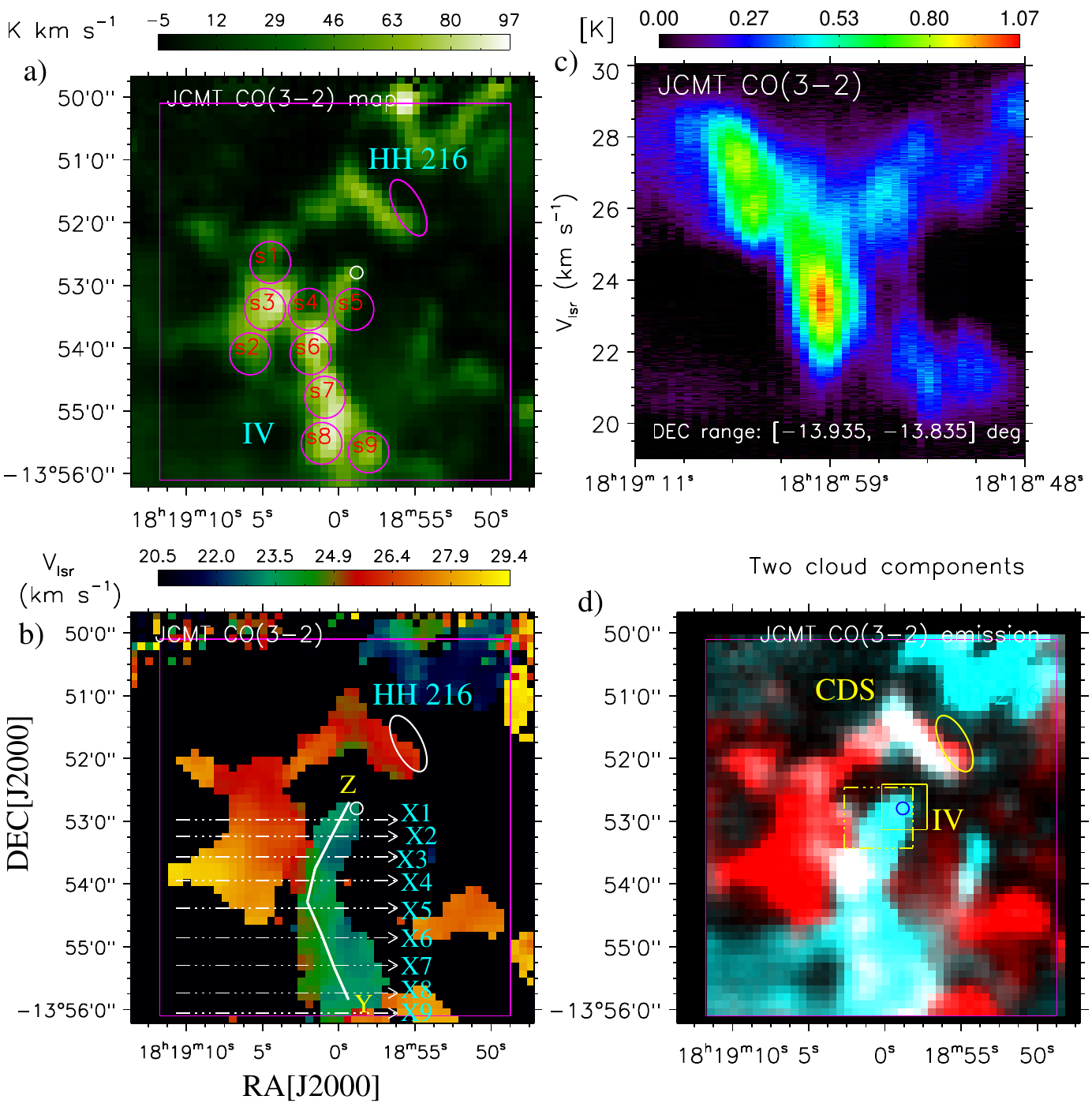}
\caption{a) JCMT CO($J$ = 3--2) integrated intensity map (at [20.514, 29.404] km s$^{-1}$) of 
an area hosting mainly the object HH~216 and the Pillar~IV (see the dot-dashed box in Figure~\ref{hfg8x}b). 
Nine circular regions (i.e., s1--s9; radii = 20 arcsec) are indicated, 
where mean velocity profiles are produced (see Figure~\ref{fgh9b}). 
b) JCMT CO($J$ = 3--2) intensity weighted velocity map. Nine arrows ``X1--X9'' and one curve ``YZ''are marked in the panel, 
where the position-velocity diagrams are produced (see Figure~\ref{fgh9c}). 
c) Position (or right ascension)-velocity diagram of the CO($J$ = 3--2) emission. 
The molecular emission is integrated over the declination range 
from $-$13.935 degrees (or $-$13$\degr$56$'$06\rlap.{$''$}0) to $-$13.835 degrees (or $-$13$\degr$50$'$06\rlap.{$''$}0) (see a solid box in Figure~\ref{fg9x}a). 
d) Spatial distribution of the JCMT CO($J$ = 3--2) gas associated with the two clouds at [25.49, 29.40] km s$^{-1}$ (in red) and 
[20.51, 25.01] km s$^{-1}$ (in turquoise). The dot-dashed box shows the area presented in Figure~\ref{fg10x}a, while the small solid box displays the area shown in Figure~\ref{fg10x}b. In panels ``a'', ``b'', and ``d'', symbols are the same as shown in Figure~\ref{hfg8x}a.}
\label{fg9x}
\end{figure*}

\begin{figure*}
\center
\includegraphics[width=0.88\textwidth]{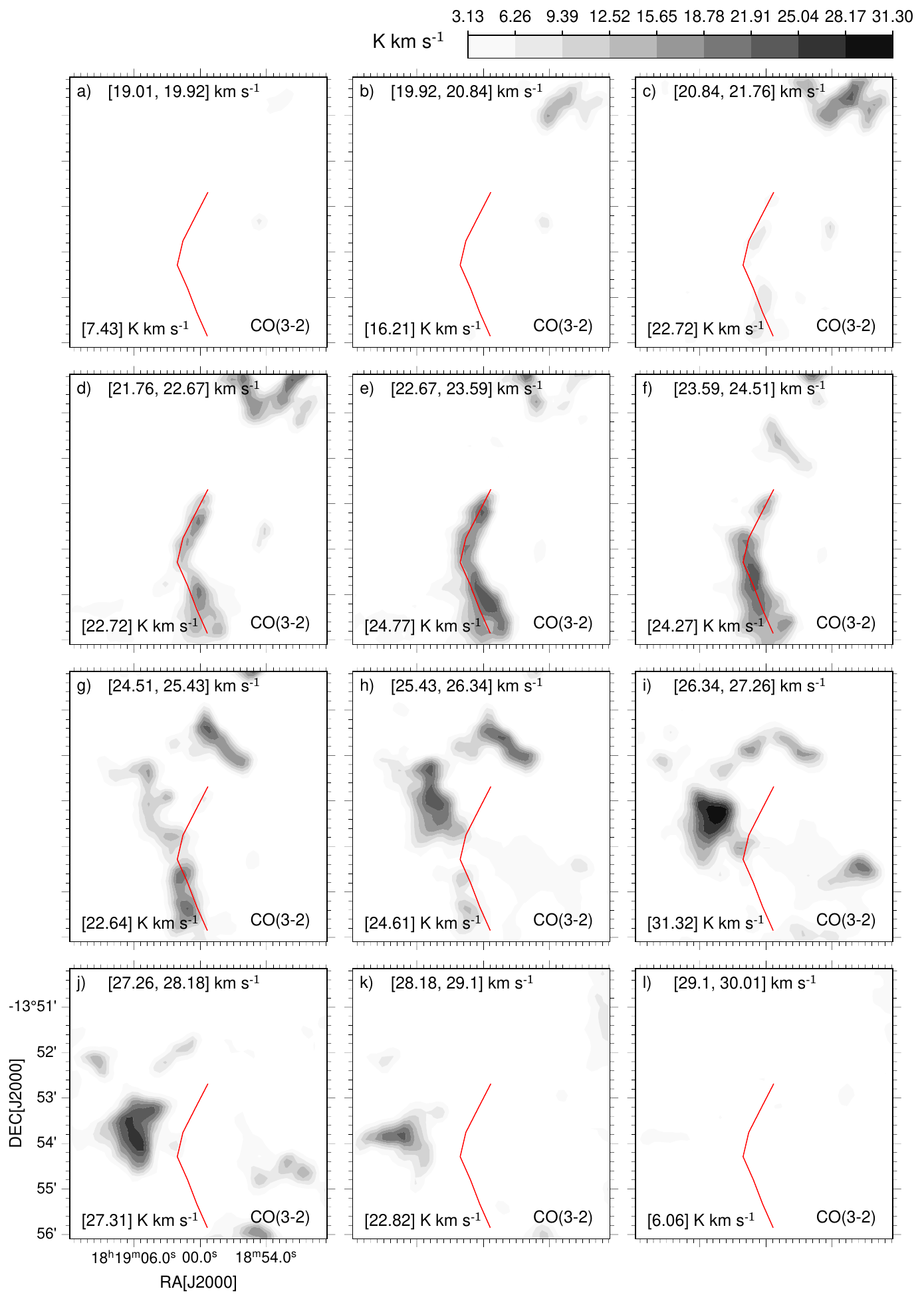}
\caption{JCMT CO($J$ = 3--2) channel maps from 19 to 30 km s$^{-1}$ with a velocity interval of 0.92 km s$^{-1}$. 
The curve (in red) is the same as shown in Figure~\ref{fg9x}b.}
\label{fgh9a}
\end{figure*}

\begin{figure*}
\center
\includegraphics[width=\textwidth]{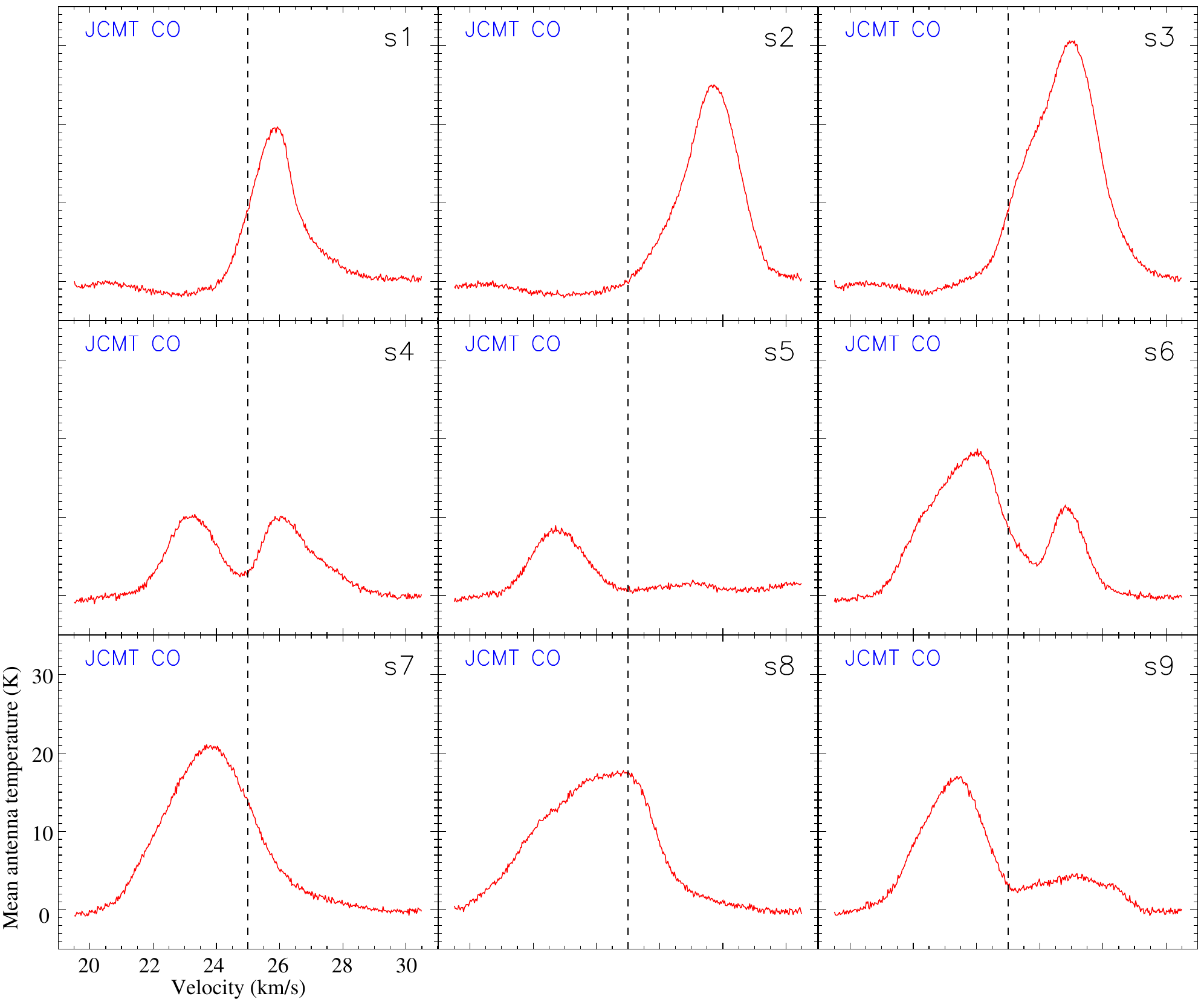}
\caption{JCMT CO($J$ = 3--2) spectra towards nine circular regions (i.e., s1--s9) marked in Figure~\ref{fg9x}a. 
In each panel, the dashed line at V$_{lsr}$ = 25 km s$^{-1}$ is highlighted.} 
\label{fgh9b}
\end{figure*}

\begin{figure*}
\center
\includegraphics[width=0.82\textwidth]{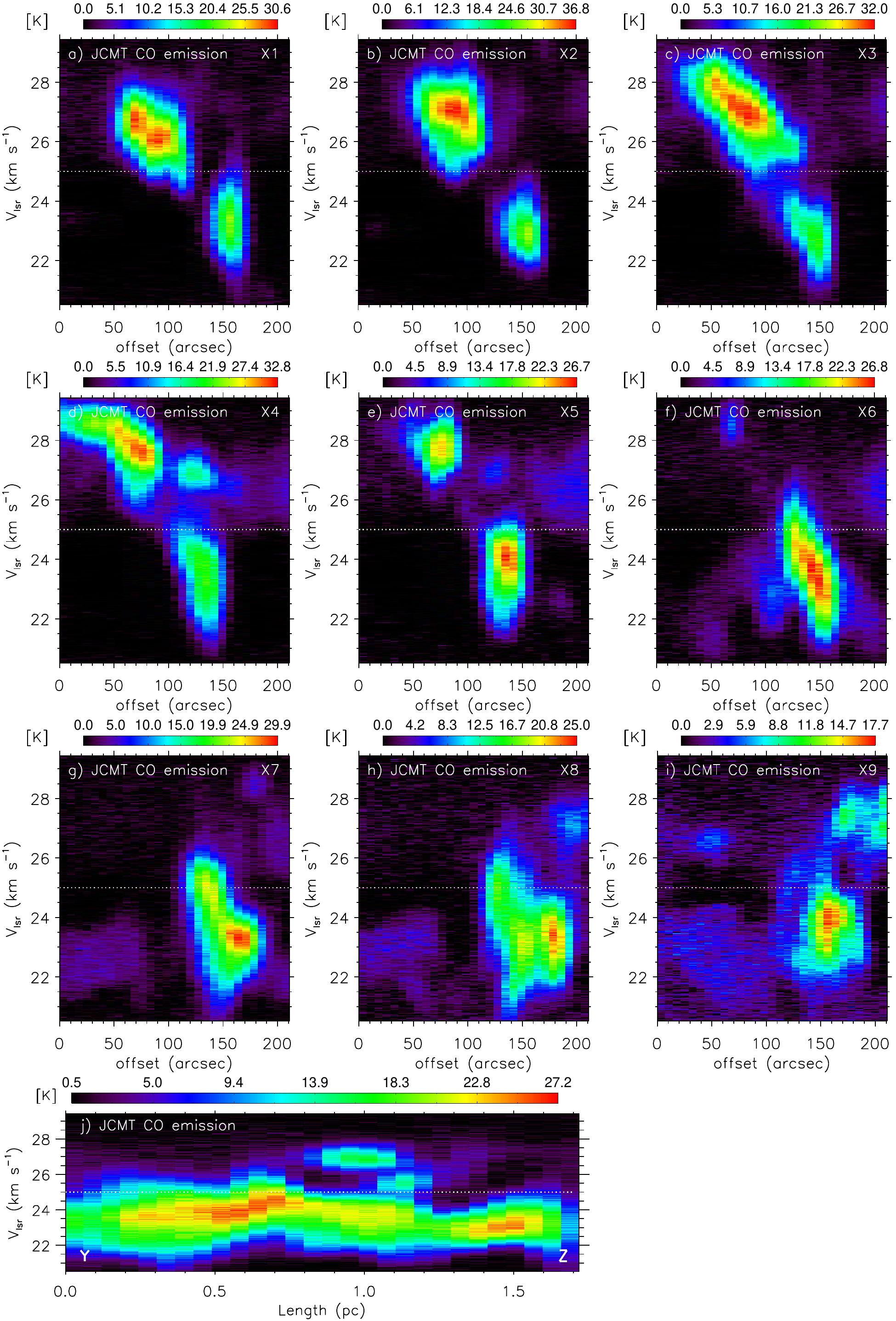}
\caption{Position-velocity diagrams of the JCMT CO($J$ = 3--2) emission along arrows 
a) ``X1''; b) ``X2''; c) ``X3''; d) ``X4''; e) ``X5''; f) ``X6''; g) ``X7''; h) ``X8''; i) ``X9''; 
and j) the curve ``YZ'' (see Figure~\ref{fg9x}b). In each panel, the dotted line at V$_{lsr}$ = 25 km s$^{-1}$ is shown.}
\label{fgh9c}
\end{figure*}

To further study the cloud components toward Pillar~IV, we have also examined the JCMT CO($J$ = 3--2) line data at [20.514, 29.404] km s$^{-1}$, which have better resolution compared to the FUGIN line data. Figures~\ref{fg9x}a and~\ref{fg9x}b display the integrated intensity and intensity weighted velocity maps of the JCMT line data, respectively. In Figures~\ref{fg9x}c, we show the position (or right ascension)-velocity diagram of the $^{12}$CO($J$ = 3--2) emission. For the position-velocity diagram, the molecular gas is integrated over the declination range from $-$13.935 to $-$13.835 degrees. 
Interestingly, the JCMT line data also support the presence of two cloud components (at [25.49, 29.40] and [20.51, 25.01] km s$^{-1}$) toward the Pillar~IV and ``CDS''. 
Figure~\ref{fg9x}d presents a two-color composite image made using the $^{12}$CO($J$ = 3--2) map at [25.49, 29.40] km s$^{-1}$ (in red) and at [20.51, 25.01] km s$^{-1}$ (in turquoise), allowing us to examine the spatial distribution of these cloud components. In the direction of the Pillar~IV and ``CDS'', overlapping areas of the two cloud components are seen in white color (see Figure~\ref{fg9x}d). To further explore these findings, more analysis of the molecular gas is presented in Figures~\ref{fgh9a},~\ref{fgh9b}, and~\ref{fgh9c}. 

Figure~\ref{fgh9a} presents the JCMT CO($J$ = 3--2) channel maps from 19 to 30 km s$^{-1}$ 
with a velocity interval of 0.92 km s$^{-1}$. On the basis of a visual appearance, we have marked an arbitrary curve 
in each panel of Figure~\ref{fgh9a}, which seems to highlight the location of the Pillar~IV (see also Figures~\ref{fg9x}b). 
From Figure~\ref{fgh9a}, we can examine the molecular gas associated with the two cloud 
components (see Figures~\ref{fgh9a}c--\ref{fgh9a}f and~\ref{fgh9a}g--\ref{fgh9a}k). 
In Figures~\ref{fgh9a}g and~\ref{fgh9a}h, we can find the common zones of the two cloud components. 
In Figure~\ref{fgh9b}, we display the spectra of the JCMT CO($J$ = 3--2) emission towards nine circular regions (s1--s9) marked in Figure~\ref{fg9x}a. In the direction of the circular regions s4 and s6, two velocity peaks are found. 
We can compare the peak velocities toward the circular regions ``s1--s3'' (around 26 km s$^{-1}$) with the peak velocities in the direction of the circular regions s5, s7, s8, and s9 (around 23 km s$^{-1}$). The study of the spectra supports the presence of two cloud components toward the Pillar~IV. 
%
Figures~\ref{fgh9c}a--\ref{fgh9c}i show position-velocity diagrams of the JCMT CO($J$ = 3--2) emission along nine arrows (see Figure~\ref{fg9x}c), while in Figure~\ref{fgh9c}j we display the position-velocity diagram along a curve ``YZ'' (see Figure~\ref{fg9x}c). 
It should be noted that this curve is chosen in the direction of Pillar~IV's length, and the arrows are selected perpendicular to Pillar~IV. 
In this direction of the Pillar~IV, these diagrams suggest the existence of two cloud components and their connection (see Figures~\ref{fgh9c}a--\ref{fgh9c}e and~\ref{fgh9c}j).   
In particular, a feature at the intermediate velocity range (i.e., [25.04, 25.46] km s$^{-1}$) is also seen between the two cloud components around 23 and 26 km s$^{-1}$ (see Figure~\ref{fgh9c}j).  

Using the JCMT CO($J$ = 3--2) line data, the integrated intensity map (at [25.04, 25.46] km s$^{-1}$) toward Pillar~IV is presented in Figure~\ref{fg10x}a. 
The selected area in Figure~\ref{fg10x}a is highlighted by the dot-dashed box in Figure~\ref{fg9x}d. 
The molecular emission traced in the intermediate velocity range is seen toward the inner parts of the Pillar~IV.  
The \emph{JWST} F470N image is overlaid with the two cloud components traced in the JCMT CO($J$ = 3--2) line data toward Pillar~IV (see Figure~\ref{fg10x}b). The driving source of the outflow seems to be located toward the common zones of the two cloud components (see the circle in Figure~\ref{fg10x}b). 

We have discussed the implication of these outcomes in more detail in Section~\ref{sec:disc}.
\section{Discussion}
\label{sec:disc}
%
The M16 H\,{\sc ii} region is an extended star-forming site ($\sim$10 pc), and is associated with the most famous astronomical object in the sky, Pillars of Creation (see Pillars I--IV in Figure~\ref{fg1x}a).
Previous works suggested that the Pillars's direction and structure are influenced by the molecular cloud's pre-existing structure, which is shaped by the radiation field of the nearby massive stars \citep[see][and references therein]{karim2023sofia,lei23}. 
Embedded protostars have been investigated toward the tips of these Pillars, where significant [C II] line and PAH emissions have been detected \citep[][see also  the \emph{JWST} F444W/F335M map in Figure~\ref{fg2x}b]{karim2023sofia}. To explain the origin of these Pillars, \citet{lei23} discussed three models, which are the instability model, cometary model, and shielding model (see their paper for more details). According to these authors, many numerical simulations have been conducted using the shielding model, which has drawn more interest in recent years. A detailed study of the Pillars’ kinematic structure is presented in \citet{karim2023sofia}. 
Note that the present work mainly focuses on the Pillar~IV (including HH~216) and the implication of the presence of molecular cloud components toward the Pillar~IV.  
\subsection{Multi-scale picture of ionized knot HH~216 and Pillar~IV}
The ionized knot, HH~216, has been considered as the bow shock of a bipolar jet \citep{meaburn82,meaburn90,andersen04,flagey20}, and 
is traced at radial velocities up to 150 km s$^{-1}$ in the [O III]$\lambda$5007 line \citep[see][]{meaburn90}. 
HH~216 is associated with the red-shifted outflow lobe, which is seen away from the Pillar~IV, while the blue-shifted outflow lobe and its driving source are traced in Pillar~IV.
According to \citet{flagey20}, the total extent of the bipolar geometry is $\sim$1.8~pc (at a distance of 2.0 kpc). \citet{andersen04} discussed the inclination of the HH flow to be at least 36 degrees 
with respect to the plane of the sky. 
The \emph{JWST} NIR and MIR images have resolved the position of the Class~I protostar that drives a bipolar outflow and is embedded in Pillar~IV (see Section~\ref{sec:phy_env}). 
The \emph{JWST} images do not detect any H$_{2}$ emission toward HH~216, however the Br$\alpha$ emission and the radio continuum emission are traced toward this object. 
\citet{anglada18} highlighted the measured centimetre radio luminosity (i.e., S$_{\nu}$d$^{2}$) values range from $\sim$100 mJy kpc$^{2}$ for massive young stars to $\sim$3 $\times$ 10$^{-3}$ mJy kpc$^{2}$ for young brown dwarfs. Using the NVAS 4.89 GHz continuum emission toward HH~216 (see Figure~\ref{fg3x}g), we have computed the total flux (S$_{\nu}$) of $\sim$22.96~mJy (size $\sim$13\rlap.{$''$}4 $\times$ 30\rlap.{$''$}2) within the contour level of 0.227 mJy beam$^{-1}$ (rms $\sim$60.2 $\mu$Jy/beam), which allows us to determine the observed centimeter radio luminosity (i.e., S$_{\nu}$d$^{2}$ (in mJy kpc$^{2}$)) to be $\sim$69.5 mJy kpc$^{2}$ at a distance ($d$) of 1.74 kpc. 

It is thought that the interaction of shocks in the associated jet with the surrounding molecular 
cloud may increase the temperature and density of the gas, which may dissociate H$_{2}$, and H may start to be ionized \citep[e.g.,][]{anglada18}. 
Hence, shocks in the jet may be responsible for the ionizing process of the radio jet or the ionized knot HH~216.

The analysis of the NVAS radio 4.89 and 8.71~GHz continuum maps favours the presence of thermal and non-thermal 
radio emission toward HH~216 (see Figure~\ref{fg4x} and also Section~\ref{sec:phy_env}). 
New low frequency radio continuum observations are needed to further confirm the non-thermal radio emission. If there exists non-thermal radio emission in HH~216 then such areas seem to be located toward its southern end, which may be linked with the wall of Pillar~II (see Section~\ref{sec:phy_env}). Such an area may be targeted for probing the physical mechanisms responsible for particle acceleration \citep[e.g.,][]{padovani15,padovani16}.

The \emph{JWST} and \emph{HST} images reveal entangled ionized structures (below 3000~AU) toward the bow shock of HH~216 (see Figure~\ref{fg5x}c), which may be located toward termination shocks.
The observed entangled structures, where at least four ionized peaks are detected in the NVAS radio 4.89~GHz continuum map, could be resultant from the interaction of shocks in the jet with its  environment. This interaction may be responsible for compression, heating, and 
variations of the magnetic field strength (and/or direction) toward HH~216 \citep[e.g.,][]{anglada18}. 
Modelling of the entangled ionized structures will be helpful, and is beyond the scope of this work. 

The \emph{JWST} images reveal new knots (i.e., HH-Na, HH-Nb, and HH-Nc) in the H$_{2}$ emission, which are located toward the northern side of Pillar~IV (see Figure~\ref{fg6x}d).
We do not find any Br$\alpha$ emission toward these knots. These observed H$_{2}$ knots support the picture of episodic accretion in the powering source of HH~216 as previously reported by \citet{flagey20}. An ionized feature (i.e., one of the parts of the jet; extent $\sim$0.16 pc) is investigated in the southern side of the protostar embedded in the central part of Pillar~IV.
This ionized feature seems to be associated with the blue-shifted lobe. We note that the NVAS radio continuum maps do not cover the area hosting Pillar~IV. Therefore, we are unable to further explore this ionized jet in this work.
In the direction of the driving source, we do not find any other point-source within a scale of 1000~AU in the \emph{JWST} images. It implies that a single source is responsible for the outflow or the episodic bursts.

Overall, the applicability of a jet/bow shock mechanism is possible as previously suggested by \citet{meaburn90} \citep[see also][]{flagey20}.
\subsection{Signatures of Cloud-Cloud Collision toward Pillar~IV}
Previously, bulk velocities around 26 and 22.5 km s$^{-1}$ were reported toward the peaks of Pillar~I and Pillar~II/III in M16, respectively \citep{pound98,white99,tremblin13}, suggesting the presence of different velocity components. In the direction of Pillars~I,~II and~III, 
\citet{karim2023sofia} presented the position-velocity diagrams of CO (see Figures 8 and 9 in their paper). 
However, these authors did not explore any position-velocity diagrams toward the Pillar~IV. Hence, there has not been done a thorough analysis of molecular line data toward Pillar~IV. In a wide spatial scale, using the NANTEN2 and FUGIN $^{12}$CO line data, \citet{nishimura21} identified two velocity components in the direction of the giant molecular cloud (GMC) associated with M16 (size $\sim$10 pc $\times$ 30 pc), which are 9.2--19.6 km s$^{-1}$ (i.e., blue-shifted component)
and 24.2--31.3 km s$^{-1}$ (i.e., red-shifted component). 
They suggested that the GMC endured numerous collision events (or Cloud-Cloud Collision (CCC)) over the course of several 10$^{6}$ years.
They also claimed that the older collision event accounts for the presence of O-type stars in the NGC~6611 cluster in M16. The age of M16 is reported to be 1.3$\pm$0.3 Myr \citep{bonatto06}. 

It has been proposed that supersonic collision between molecular clouds produces a shock-compressed interface with amplified magnetic field \citep[e.g.,][and references therein]{habe92,anathpindika10,inoue13,haworth15a,haworth15b,torii17,balfour17,bisbas17}, where massive stars and clusters of YSOs can be formed. 
In favour of the proposed collision event, we find several observational works in the literature \citep[e.g.,][]{torii11,torii15,torii17,fukui14,fukui18,fukui21,dhanya21,maity22,maity23}.
In the collision event, we expect at least two cloud components and their connection in both spatial and velocity space. In such process, a bridge feature, which is a low-intensity feature connecting two cloud components with an intermediate velocity range in position-velocity diagrams, is anticipated \citep[e.g.,][]{haworth15b,dewangan17s235,dewangan18b,Kohno18,Priestley21}.  
%
The bridge feature has been suggested to hint the presence of the turbulent motion of the gas enhanced by the collision and the shocked interface layer \citep{haworth15a,haworth15b,torii17}. 
%
Moreover, we expect a complementary distribution in the CCC event, aligning with the spatial correlation of ``key/intensity-enhancement'' and ``cavity/keyhole/intensity-depression'' characteristics \citep[e.g.,][]{fukui18,dewangan18N36,Enokiya21}. 

In Section~\ref{sec:coem}, the analysis of the FUGIN and JCMT molecular line data has revealed similar observational signposts as expected 
in the collision of molecular clouds.
In the direction of Pillar~IV, two cloud components around 23 and 26 km $^{-1}$ and the complementary distribution of these components are investigated without any spatial displacement. 
A spatial match of the component around 23 km $^{-1}$ (i.e., ``key/intensity-enhancement'') and the component around 26 km s$^{-1}$ (i.e., ``cavity/keyhole/intensity-depression'') is investigated (see Section~\ref{sec:coem} for more details). Additionally, we also find almost a V-like velocity structure and a possible bridge-like feature in the position-velocity diagrams of the JCMT CO($J$ = 3--2) emission \citep[see also][]{maity22}. All these observational features favour that the gas at different velocities might have collided, and does not appear to be simply superposed along the line of sight. 

%
%
\subsubsection{Estimation of collision time-scale}
\label{sec:ccc_timescale}
\begin{figure*}
\center
\includegraphics[width=0.7\textwidth]{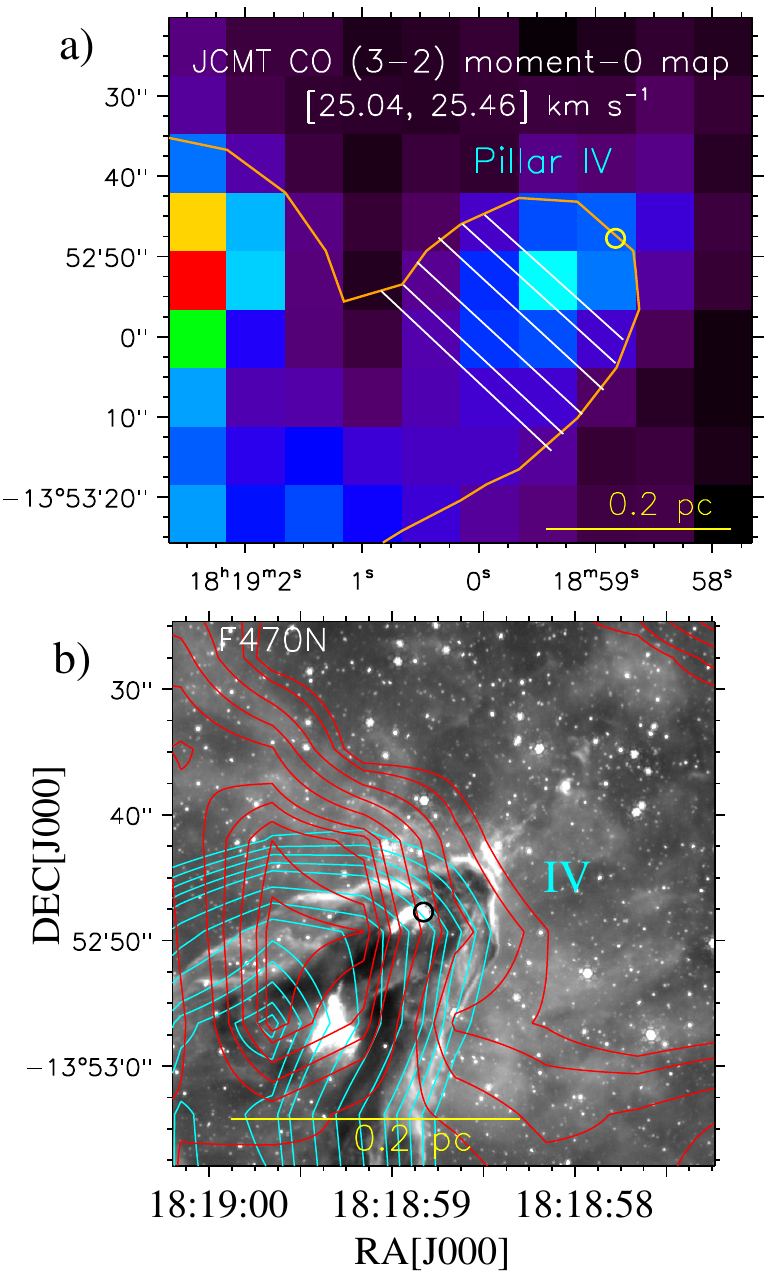}
\caption{a) JCMT CO($J$ = 3--2) integrated intensity map for the intermediate velocity range (i.e., [25.04, 25.46] km s$^{-1}$) toward Pillar~IV for an area highlighted by the dot-dashed box in Figure~\ref{fg9x}d. The orange contour is at 6$\sigma$ level, where, 1$\sigma \sim$0.1 K km s$^{-1}$. Six lines (in white) used to determine the half-width/radius of the shock compressed layer are marked toward Pillar~IV. b) Distribution of the JCMT CO($J$ = 3--2) emission at [25.49, 29.40] and [20.51, 25.01] km s$^{-1}$ toward Pillar~IV (see the solid box in Figure~\ref{fg9x}d). The cyan contours at [20.51, 25.01] km s$^{-1}$ are 56.14 $\times$ (0.2, 0.25, 0.3, 0.35, 0.4, 0.5, 0.6, 0.7, 0.8, 0.85, 0.9, 0.95, and 0.98) K km s$^{-1}$.The red contours at [25.49, 29.40] km s$^{-1}$ are 5.59 $\times$ (0.35, 0.4, 0.5, 0.6, 0.7, 0.8, 0.85, 0.9, 0.95, and 0.98) K km s$^{-1}$. The circle represents the location of the driving source of the previously reported outflow and a scale bar corresponding to 0.2 kpc at a distance of 1.74 kpc is drawn in each panel.}
\label{fg10x}
\end{figure*}

For a collisional velocity of $v_{\rm col}$ between molecular clouds with an initial density $n_{1}$ and the magnetic field component perpendicular to the axis of collision $B_{1}$, the shock compression ratio is given by \citet{fukui21} as follows,
\begin{equation}\label{eqMA}
\frac{n_2}{n_1} = \frac{B_2}{B_1}\simeq17\,\left(  \frac{v_{\rm col}}{10\mbox{ km s}^{-1}} \right)\,\left(  \frac{B_{1}}{10\,\mu\mbox{G}} \right)^{-1}\,\left(  \frac{n_{1}}{300\mbox{ cm}^{-3}} \right)^{1/2}
\end{equation}
Here, $n_{2}$ and $B_{2}$ correspond to the final density and magnetic field component perpendicular to the axis of collision, respectively. As discussed in \citet{maity23}, the requirement of zero spatial shift for the complementary distribution hints that the angle between the line of observation and axis of collision $\sim0$ degree \citep[see also][]{fukui18,fukui21}. 
Therefore, in our current study, the value of $v_{\rm col}$ is $\sim \frac{3.0}{\cos 0^{\circ}}$ km s$^{-1}$ = 3 km s$^{-1}$. 
By employing the Chandrasekhar–Fermi methodology \citep{Chandrasekhar_1953ApJ}, \citet{Pattle_2018ApJ} estimated the plane of sky magnetic field strength for the ``Pillars of Creation'', which ranges from 170 to 320 $\mu$G. 
Hence, it is safe to choose $B_2$ = 200--300 $\mu$G (after collision) and $B_1$ = 10 $\mu$G \citep[before collision;][]{Crutcher_2012}. Now, following Equation~\ref{eqMA} we find $\frac{n_2}{n_1} =$ 20--30, and $n_1\sim$ 4.6--10 $\times$10$^{3}$ cm$^{-3}$. After the initial collision between two clouds, the onset of gravitational instability can be estimated using the following equation \citep[e.g.,][]{Whitworth_1994,fukui21}:
\begin{equation}\label{eq:W94}
t_{\rm start} \sim 0.5\,\frac{\left( \frac{c_{\rm s}}{0.2\mbox{ km s}^{-1}} \right)^{1/2}}{\left( \frac{n_{1}}{300\mbox{ cm}^{-3}} \right)^{1/2}\left( \frac{v_{\rm col}}{10\mbox{ km s}^{-1}} \right)^{1/2}} \mbox{ Myr}.
\end{equation}
In this equation, $c_\mathrm{s}$ represents the sound speed in the molecular cloud before the collision. Using $c_\mathrm{s}$ = 0.2 km s$^{-1}$, $v_{\rm col}$ = 3 km s$^{-1}$, and $n_1\sim$ 4.6-10 $\times$10$^{3}$ cm$^{-3}$ in Equation~\ref{eq:W94}, we find that $t_{\rm start}$ falls within the range of 0.15 to 0.23 Myr. Drawing from the approach introduced by \citet{henshaw13}, the collision timescale ($t_{\rm col}$) can be determined using the following equation:
\begin{equation}\label{tcollision}
t_{\rm col} \sim {2.0\, \bigg(\frac{R_f}{0.5\,{\rm pc}} \bigg) \bigg(\frac{v_{\rm col}}{5{\rm\,km\,s^{-1}}}\bigg)^{-1}\bigg(\frac{n_{2}/n_{1}}{10}\bigg)\,{\rm Myr}},
\end{equation} 
where $R_f$ is the radius or half-width of the higher density region created through the collision event. In the direction of Pillar~IV, we determined the value of $R_f$ using the JCMT CO($J$ = 3--2) integrated intensity map for the intermediate velocity range (i.e., [25.04, 25.46]~km s$^{-1}$), which corresponds to the bridge feature. 
The value of $R_f$ is determined to be $\sim$0.113 pc, which is calculated as half of the average length of the lines marked on Pillar~IV (see Figure~\ref{fg10x}a). 
With our estimated values for $\frac{n_{2}}{n_{1}}$, $v_{\rm col}$, and $R_f$, the resulting collision timescale, $t_{\rm col}$, is computed in a range of [1.5, 2.3] Myr. 
Notably, this timescale roughly equals or exceeds the combined values of the previously reported age of M16 and the timescale for the onset of gravitational instability following the collision event.  

The \emph{JWST} images show the presence of an isolated Class~I protostar in the Pillar~IV, and the mean age of a Class~I protostar has been reported to be $\sim$0.44 Myr \citep{evans09}. As highlighted earlier, we cannot ignore the formation of the Pillar~IV by radiative compression/ablation similar to Pillars I, II, and 
III \citep[e.g.,][]{karim2023sofia}. In this context, the embedded YSO driving HH~216 may be formed by radiative implosion or the feedback of massive stars 
\citep[see][for more details]{karim2023sofia}. However, our results also suggest that the interaction of molecular cloud components might have influenced star formation activity in Pillar~IV.  
%

%

%

\section{Summary and Conclusions}
\label{sec:conc}
The present work uses multi-scale and multi-wavelength data sets to probe ongoing physical processes 
toward the Pillar~IV and the ionized knot HH~216 in Eagle nebula (M16; d $\sim$1.74 kpc). A bipolar outflow (extent $\sim$1.6 pc at a distance of 1.74 kpc)  has been reported, and the outflow-driving source is identified as a Class~I protostar embedded in Pillar~IV. 
HH~216 has been recognized as the bow shock of the red-shifted outflow lobe, and is associated with the radio continuum emission. This protostar is observed as a single, isolated object (below 1000~AU) in the \emph{JWST} NIR and MIR 
images (resolution $\sim$0\rlap.{$''$}07 -- 0\rlap.{$''$}7). 
Using the \emph{JWST} F444W/F335M image, the outer boundary of Pillar~IV is depicted with the 3.3 $\mu$m PAH emission.  
HH~216 is found to be associated with the 4.05 $\mu$m Br$\alpha$ emission in the \emph{JWST} F444W/F335M image and the radio 4.89 and 8.71 GHz continuum emission. However, it is not detected with the 4.693 $\mu$m H$_{2}$ emission in the \emph{JWST} F470N$-$F444W image. 
Using the NVAS radio 4.89 and 8.71 GHz continuum maps, we have produced the radio spectral index, 
indicating that HH~216 appears to be associated with both thermal and non-thermal radio emission.
High resolution \emph{JWST} and HST images uncover entangled ionized structures (below 3000~AU) toward HH~216, which seem to be positioned near termination shocks. 
The \emph{JWST} F470N$-$F444W image has allowed us to identify new knots associated with the 4.693 $\mu$m H$_{2}$ emission, which are distributed toward the northern side of Pillar~IV. This particular finding favours the previously proposed hypothesis of episodic accretion in the powering source of HH~216. 
On the southern side of the driving source, an ionized feature (extent $\sim$0.16 pc) or one of the parts of the jet is investigated. 
The ionized emission can be an outcome of the interplay between the shocks in the jet with its surroundings.

Using the JCMT CO($J$ = 3--2), FUGIN $^{12}$CO($J$ = 1--0), and FUGIN $^{13}$CO($J$ = 1--0) emission, observational signposts of CCC toward the Pillar~IV are probed, which are the connection of two cloud components (around 23 and 26 km $^{-1}$) in both physical and velocity space, a bridge-like feature, a complementary distribution of cloud components, and a possible V-like velocity structure. 
Overall, our findings imply that star formation activity in Pillar~IV may have been impacted by the interaction of molecular cloud components between 23 and 26 km s$^{-1}$, which may have happened $\sim$1.5-2.3 Myr ago. 
\section*{Acknowledgments}
We thank the anonymous reviewer for the constructive comments and suggestions.   
The research work at Physical Research Laboratory is funded by the Department of Space, Government of India. 
This work is based [in part] on observations made with the {\it Spitzer} Space Telescope, which is operated by the Jet Propulsion Laboratory, California Institute of Technology under a contract with NASA. 
This work is based on observations made with the NASA/ESA/CSA James Webb Space Telescope.
The data were obtained from the Mikulski Archive for Space Telescopes at the Space Telescope Science Institute, which is operated
by the Association of Universities for Research in Astronomy, Inc.,
under NASA contract NAS 5-03127 for \emph{JWST}. These observations
are associated with the program \#2739\footnote[1]{https://archive.stsci.edu/doi/resolve/resolve.html?doi=10.17909/r4d5-d269}.  
The James Clerk Maxwell Telescope is operated by the East Asian Observatory on behalf of The National Astronomical 
Observatory of Japan; Academia Sinica Institute of Astronomy and Astrophysics; the Korea Astronomy and Space Science Institute; 
the National Astronomical Research Institute of Thailand; Center for Astronomical Mega-Science (as well as the National Key R\&D 
Program of China with No. 2017YFA0402700). Additional funding support is provided by the Science and Technology Facilities 
Council of the United Kingdom and participating universities and organizations in the United Kingdom and Canada. 
Additional funds for the construction of SCUBA-2 were provided by the Canada Foundation for Innovation. 
This research has made use of the NASA/IPAC Infrared Science Archive, which is funded by the National Aeronautics and Space Administration and operated by the California Institute of Technology. The NVAS image was produced as part of the NRAO VLA Archive Survey, (c) AUI/NRAO. This research is based [in part] on observations made with the NASA/ESA Hubble Space Telescope obtained from the Space Telescope Science Institute (Proposal ID: 13926; PI: Zolt Levay) , which is operated by the Association of Universities for Research in Astronomy, Inc., under NASA contract NAS 5–26555. This publication makes use of data from FUGIN, FOREST Unbiased Galactic 
plane Imaging survey with the Nobeyama 45-m telescope, a legacy project in the Nobeyama 45-m radio telescope. This research made use of {\it Astropy}\footnote[1]{http://www.astropy.org}, a community-developed core Python package for Astronomy \citep{astropy13,astropy18}. For figures, we have used {\it matplotlib} \citep{Hunter_2007} and IDL software.  
\subsection*{Data availability}
The NVAS radio continuum data underlying this article are available from the publicly accessible server\footnote[2]{http://www.vla.nrao.edu/astro/nvas/}.
The FUGIN molecular line data underlying this article is available from the website of the VizieR Service\footnote[3]{https://nro-fugin.github.io/release/}.
The SCUBA-2 850 $\mu$m map and HARP CO line data underlying this article are available from the publicly accessible JCMT science archive\footnote[4]{https://www.cadc-ccda.hia-iha.nrc-cnrc.gc.ca/en/jcmt/}.
The {\it Herschel} column density and temperature maps underlying this article are available from the publicly accessible website\footnote[5]{http://www.astro.cardiff.ac.uk/research/ViaLactea/}.
The {\it Spitzer} images underlying this article are available from the publicly accessible NASA/IPAC infrared science archive\footnote[6]{https://irsa.ipac.caltech.edu/frontpage/}.
The \emph{JWST} images underlying this article are available from the publicly accessible MAST archive\footnote[7]{https://archive.stsci.edu/missions-and-data/jwst}.
The HST F657N image underlying this article is available from the publicly accessible JCMT science archive\footnote[8]{https://archive.stsci.edu/missions-and-data/hst/}.
%


\bibliographystyle{mnras}
\bibliography{reference} 

\begin{thebibliography}{}
\makeatletter
\relax
\def\mn@urlcharsother{\let\do\@makeother \do\$\do\&\do\#\do\^\do\_\do\%\do\~}
\def\mn@doi{\begingroup\mn@urlcharsother \@ifnextchar [ {\mn@doi@}
  {\mn@doi@[]}}
\def\mn@doi@[#1]#2{\def\@tempa{#1}\ifx\@tempa\@empty \href
  {http://dx.doi.org/#2} {doi:#2}\else \href {http://dx.doi.org/#2} {#1}\fi
  \endgroup}
\def\mn@eprint#1#2{\mn@eprint@#1:#2::\@nil}
\def\mn@eprint@arXiv#1{\href {http://arxiv.org/abs/#1} {{\tt arXiv:#1}}}
\def\mn@eprint@dblp#1{\href {http://dblp.uni-trier.de/rec/bibtex/#1.xml}
  {dblp:#1}}
\def\mn@eprint@#1:#2:#3:#4\@nil{\def\@tempa {#1}\def\@tempb {#2}\def\@tempc
  {#3}\ifx \@tempc \@empty \let \@tempc \@tempb \let \@tempb \@tempa \fi \ifx
  \@tempb \@empty \def\@tempb {arXiv}\fi \@ifundefined
  {mn@eprint@\@tempb}{\@tempb:\@tempc}{\expandafter \expandafter \csname
  mn@eprint@\@tempb\endcsname \expandafter{\@tempc}}}

\bibitem[\protect\citeauthoryear{{Anathpindika}}{{Anathpindika}}{2010}]{anathpindika10}
{Anathpindika} S.~V.,  2010, \mn@doi [\mnras]
  {10.1111/j.1365-2966.2010.16541.x}, \href
  {https://ui.adsabs.harvard.edu/abs/2010MNRAS.405.1431A} {405, 1431}

\bibitem[\protect\citeauthoryear{{Andersen}, {Knude}, {Reipurth}, {Castets},
  {Nyman}, {McCaughrean}  \& {Heathcote}}{{Andersen} et~al.}{2004}]{andersen04}
{Andersen} M.,  {Knude} J.,  {Reipurth} B.,  {Castets} A.,  {Nyman}
  L.~{\r{A}}.,  {McCaughrean} M.~J.,   {Heathcote} S.,  2004, \mn@doi [\aap]
  {10.1051/0004-6361:20031535}, \href
  {https://ui.adsabs.harvard.edu/abs/2004A&A...414..969A} {414, 969}

\bibitem[\protect\citeauthoryear{{Anglada}, {Rodr{\'\i}guez}  \&
  {Carrasco-Gonz{\'a}lez}}{{Anglada} et~al.}{2018}]{anglada18}
{Anglada} G.,  {Rodr{\'\i}guez} L.~F.,   {Carrasco-Gonz{\'a}lez} C.,  2018,
  \mn@doi [\aapr] {10.1007/s00159-018-0107-z}, \href
  {https://ui.adsabs.harvard.edu/abs/2018A&ARv..26....3A} {26, 3}

\bibitem[\protect\citeauthoryear{{Astropy Collaboration} et~al.,}{{Astropy
  Collaboration} et~al.}{2013}]{astropy13}
{Astropy Collaboration} et~al., 2013, \mn@doi [\aap]
  {10.1051/0004-6361/201322068}, \href
  {https://ui.adsabs.harvard.edu/abs/2013A&A...558A..33A} {558, A33}

\bibitem[\protect\citeauthoryear{{Astropy Collaboration} et~al.,}{{Astropy
  Collaboration} et~al.}{2018}]{astropy18}
{Astropy Collaboration} et~al., 2018, \mn@doi [\aj] {10.3847/1538-3881/aabc4f},
  \href {https://ui.adsabs.harvard.edu/abs/2018AJ....156..123A} {156, 123}

\bibitem[\protect\citeauthoryear{{Balfour}, {Whitworth}  \& {Hubber}}{{Balfour}
  et~al.}{2017}]{balfour17}
{Balfour} S.~K.,  {Whitworth} A.~P.,   {Hubber} D.~A.,  2017, \mn@doi [\mnras]
  {10.1093/mnras/stw2956}, \href
  {https://ui.adsabs.harvard.edu/abs/2017MNRAS.465.3483B} {465, 3483}

\bibitem[\protect\citeauthoryear{{Beichman}, {Rieke}, {Eisenstein}, {Greene},
  {Krist}, {McCarthy}, {Meyer}  \& {Stansberry}}{{Beichman}
  et~al.}{2012}]{2012SPIE.8442E..2NB}
{Beichman} C.~A.,  {Rieke} M.,  {Eisenstein} D.,  {Greene} T.~P.,  {Krist} J.,
  {McCarthy} D.,  {Meyer} M.,   {Stansberry} J.,  2012, in {Clampin} M.~C.,
  {Fazio} G.~G.,  {MacEwen} H.~A.,   {Oschmann} Jacobus~M. J.,  eds,  Society
  of Photo-Optical Instrumentation Engineers (SPIE) Conference Series Vol.
  8442, Space Telescopes and Instrumentation 2012: Optical, Infrared, and
  Millimeter Wave. p. 84422N, \mn@doi{10.1117/12.925447}

\bibitem[\protect\citeauthoryear{{Benjamin} et~al.,}{{Benjamin}
  et~al.}{2003}]{benjamin03}
{Benjamin} R.~A.,  et~al., 2003, \mn@doi [\pasp] {10.1086/376696}, \href
  {https://ui.adsabs.harvard.edu/abs/2003PASP..115..953B} {115, 953}

\bibitem[\protect\citeauthoryear{{Bisbas}, {Tanaka}, {Tan}, {Wu}  \&
  {Nakamura}}{{Bisbas} et~al.}{2017}]{bisbas17}
{Bisbas} T.~G.,  {Tanaka} K. E.~I.,  {Tan} J.~C.,  {Wu} B.,   {Nakamura} F.,
  2017, \mn@doi [\apj] {10.3847/1538-4357/aa94c5}, \href
  {https://ui.adsabs.harvard.edu/abs/2017ApJ...850...23B} {850, 23}

\bibitem[\protect\citeauthoryear{{Bonatto}, {Santos}  \& {Bica}}{{Bonatto}
  et~al.}{2006}]{bonatto06}
{Bonatto} C.,  {Santos} J.~F.~C. J.,   {Bica} E.,  2006, \mn@doi [\aap]
  {10.1051/0004-6361:20052793}, \href
  {https://ui.adsabs.harvard.edu/abs/2006A&A...445..567B} {445, 567}

\bibitem[\protect\citeauthoryear{{Buckle} et~al.,}{{Buckle}
  et~al.}{2009}]{buckle09}
{Buckle} J.~V.,  et~al., 2009, \mn@doi [\mnras]
  {10.1111/j.1365-2966.2009.15347.x}, \href
  {https://ui.adsabs.harvard.edu/abs/2009MNRAS.399.1026B} {399, 1026}

\bibitem[\protect\citeauthoryear{{Chandrasekhar} \& {Fermi}}{{Chandrasekhar} \&
  {Fermi}}{1953}]{Chandrasekhar_1953ApJ}
{Chandrasekhar} S.,  {Fermi} E.,  1953, \mn@doi [\apj] {10.1086/145731}, \href
  {https://ui.adsabs.harvard.edu/abs/1953ApJ...118..113C} {118, 113}

\bibitem[\protect\citeauthoryear{{Crossley}, {Sjouwerman}, {Fomalont}  \&
  {Radziwill}}{{Crossley} et~al.}{2007}]{crossley07}
{Crossley} J.~H.,  {Sjouwerman} L.~O.,  {Fomalont} E.~B.,   {Radziwill} N.~M.,
  2007, in American Astronomical Society Meeting Abstracts. p. 132.03

\bibitem[\protect\citeauthoryear{{Crutcher}}{{Crutcher}}{2012}]{Crutcher_2012}
{Crutcher} R.~M.,  2012, \mn@doi [\araa] {10.1146/annurev-astro-081811-125514},
  \href {https://ui.adsabs.harvard.edu/abs/2012ARA&A..50...29C} {50, 29}

\bibitem[\protect\citeauthoryear{{Dewangan} \& {Ojha}}{{Dewangan} \&
  {Ojha}}{2017}]{dewangan17s235}
{Dewangan} L.~K.,  {Ojha} D.~K.,  2017, \mn@doi [\apj]
  {10.3847/1538-4357/aa8e00}, \href
  {https://ui.adsabs.harvard.edu/abs/2017ApJ...849...65D} {849, 65}

\bibitem[\protect\citeauthoryear{{Dewangan}, {Ojha}, {Zinchenko}, {Janardhan}
  \& {Luna}}{{Dewangan} et~al.}{2017}]{dewangan17a}
{Dewangan} L.~K.,  {Ojha} D.~K.,  {Zinchenko} I.,  {Janardhan} P.,   {Luna} A.,
   2017, \mn@doi [\apj] {10.3847/1538-4357/834/1/22}, \href
  {https://ui.adsabs.harvard.edu/abs/2017ApJ...834...22D} {834, 22}

\bibitem[\protect\citeauthoryear{{Dewangan}, {Ojha}, {Zinchenko}  \&
  {Baug}}{{Dewangan} et~al.}{2018a}]{dewangan18b}
{Dewangan} L.~K.,  {Ojha} D.~K.,  {Zinchenko} I.,   {Baug} T.,  2018a, \mn@doi
  [\apj] {10.3847/1538-4357/aac6bb}, \href
  {https://ui.adsabs.harvard.edu/abs/2018ApJ...861...19D} {861, 19}

\bibitem[\protect\citeauthoryear{{Dewangan}, {Dhanya}, {Ojha}  \&
  {Zinchenko}}{{Dewangan} et~al.}{2018b}]{dewangan18N36}
{Dewangan} L.~K.,  {Dhanya} J.~S.,  {Ojha} D.~K.,   {Zinchenko} I.,  2018b,
  \mn@doi [\apj] {10.3847/1538-4357/aadfe3}, \href
  {https://ui.adsabs.harvard.edu/abs/2018ApJ...866...20D} {866, 20}

\bibitem[\protect\citeauthoryear{{Dewangan}, {Ojha}, {Sharma}, {Palacio},
  {Bhadari}  \& {Das}}{{Dewangan} et~al.}{2020}]{dewangan20g}
{Dewangan} L.~K.,  {Ojha} D.~K.,  {Sharma} S.,  {Palacio} S.~d.,  {Bhadari}
  N.~K.,   {Das} A.,  2020, \mn@doi [\apj] {10.3847/1538-4357/abb827}, \href
  {https://ui.adsabs.harvard.edu/abs/2020ApJ...903...13D} {903, 13}

\bibitem[\protect\citeauthoryear{{Dewangan}, {Maity}, {Mayya}, {Bhadari},
  {Bhattacharyya}, {Sharma}  \& {Banerjee}}{{Dewangan}
  et~al.}{2023}]{dewangan2023new}
{Dewangan} L.~K.,  {Maity} A.~K.,  {Mayya} Y.~D.,  {Bhadari} N.~K.,
  {Bhattacharyya} S.,  {Sharma} S.,   {Banerjee} G.,  2023, \mn@doi [arXiv
  e-prints] {10.48550/arXiv.2309.13351}, \href
  {https://ui.adsabs.harvard.edu/abs/2023arXiv230913351D} {p. arXiv:2309.13351}

\bibitem[\protect\citeauthoryear{{Dhanya}, {Dewangan}, {Ojha}  \&
  {Mandal}}{{Dhanya} et~al.}{2021}]{dhanya21}
{Dhanya} J.~S.,  {Dewangan} L.~K.,  {Ojha} D.~K.,   {Mandal} S.,  2021, \mn@doi
  [\pasj] {10.1093/pasj/psz137}, \href
  {https://ui.adsabs.harvard.edu/abs/2021PASJ...73S.355D} {73, S355}

\bibitem[\protect\citeauthoryear{{Enokiya}, {Torii}  \& {Fukui}}{{Enokiya}
  et~al.}{2021}]{Enokiya21}
{Enokiya} R.,  {Torii} K.,   {Fukui} Y.,  2021, \mn@doi [\pasj]
  {10.1093/pasj/psz119}, \href
  {https://ui.adsabs.harvard.edu/abs/2021PASJ...73S..75E} {73, S75}

\bibitem[\protect\citeauthoryear{{Evans} Neal~J. et~al.,}{{Evans}
  et~al.}{2009}]{evans09}
{Evans} Neal~J. I.,  et~al., 2009, \mn@doi [\apjs]
  {10.1088/0067-0049/181/2/321}, \href
  {https://ui.adsabs.harvard.edu/abs/2009ApJS..181..321E} {181, 321}

\bibitem[\protect\citeauthoryear{{Flagey}, {McLeod}, {Aguilar}  \&
  {Prunet}}{{Flagey} et~al.}{2020}]{flagey20}
{Flagey} N.,  {McLeod} A.~F.,  {Aguilar} L.,   {Prunet} S.,  2020, \mn@doi
  [\aap] {10.1051/0004-6361/201833690}, \href
  {https://ui.adsabs.harvard.edu/abs/2020A&A...635A.111F} {635, A111}

\bibitem[\protect\citeauthoryear{{Fukui} et~al.,}{{Fukui}
  et~al.}{2014}]{fukui14}
{Fukui} Y.,  et~al., 2014, \mn@doi [\apj] {10.1088/0004-637X/780/1/36}, \href
  {https://ui.adsabs.harvard.edu/abs/2014ApJ...780...36F} {780, 36}

\bibitem[\protect\citeauthoryear{{Fukui} et~al.,}{{Fukui}
  et~al.}{2018}]{fukui18}
{Fukui} Y.,  et~al., 2018, \mn@doi [\apj] {10.3847/1538-4357/aac217}, \href
  {https://ui.adsabs.harvard.edu/abs/2018ApJ...859..166F} {859, 166}

\bibitem[\protect\citeauthoryear{{Fukui}, {Habe}, {Inoue}, {Enokiya}  \&
  {Tachihara}}{{Fukui} et~al.}{2021}]{fukui21}
{Fukui} Y.,  {Habe} A.,  {Inoue} T.,  {Enokiya} R.,   {Tachihara} K.,  2021,
  \mn@doi [\pasj] {10.1093/pasj/psaa103}, \href
  {https://ui.adsabs.harvard.edu/abs/2021PASJ...73S...1F} {73, S1}

\bibitem[\protect\citeauthoryear{{GLIMPSE Team}}{{GLIMPSE
  Team}}{2020}]{glimpse1}
{GLIMPSE Team} 2020, GLIMPSE I Archive, \mn@doi{10.26131/IRSA210}, \url
  {https://catcopy.ipac.caltech.edu/dois/doi.php?id=10.26131/IRSA210}

\bibitem[\protect\citeauthoryear{{Getman}, {Feigelson}, {Garmire}, {Broos}  \&
  {Wang}}{{Getman} et~al.}{2007}]{getman07}
{Getman} K.~V.,  {Feigelson} E.~D.,  {Garmire} G.,  {Broos} P.,   {Wang} J.,
  2007, \mn@doi [\apj] {10.1086/509112}, \href
  {https://ui.adsabs.harvard.edu/abs/2007ApJ...654..316G} {654, 316}

\bibitem[\protect\citeauthoryear{{Habe} \& {Ohta}}{{Habe} \&
  {Ohta}}{1992}]{habe92}
{Habe} A.,  {Ohta} K.,  1992, \pasj, \href
  {https://ui.adsabs.harvard.edu/abs/1992PASJ...44..203H} {44, 203}

\bibitem[\protect\citeauthoryear{{Hartmann}, {Megeath}, {Allen}, {Luhman},
  {Calvet}, {D'Alessio}, {Franco-Hernandez}  \& {Fazio}}{{Hartmann}
  et~al.}{2005}]{hartmann05}
{Hartmann} L.,  {Megeath} S.~T.,  {Allen} L.,  {Luhman} K.,  {Calvet} N.,
  {D'Alessio} P.,  {Franco-Hernandez} R.,   {Fazio} G.,  2005, \mn@doi [\apj]
  {10.1086/431472}, \href
  {https://ui.adsabs.harvard.edu/abs/2005ApJ...629..881H} {629, 881}

\bibitem[\protect\citeauthoryear{{Haworth} et~al.,}{{Haworth}
  et~al.}{2015a}]{haworth15a}
{Haworth} T.~J.,  et~al., 2015a, \mn@doi [\mnras] {10.1093/mnras/stv639}, \href
  {https://ui.adsabs.harvard.edu/abs/2015MNRAS.450...10H} {450, 10}

\bibitem[\protect\citeauthoryear{{Haworth}, {Shima}, {Tasker}, {Fukui},
  {Torii}, {Dale}, {Takahira}  \& {Habe}}{{Haworth} et~al.}{2015b}]{haworth15b}
{Haworth} T.~J.,  {Shima} K.,  {Tasker} E.~J.,  {Fukui} Y.,  {Torii} K.,
  {Dale} J.~E.,  {Takahira} K.,   {Habe} A.,  2015b, \mn@doi [\mnras]
  {10.1093/mnras/stv2068}, \href
  {https://ui.adsabs.harvard.edu/abs/2015MNRAS.454.1634H} {454, 1634}

\bibitem[\protect\citeauthoryear{{Healy}, {Hester}  \& {Claussen}}{{Healy}
  et~al.}{2004}]{healy04}
{Healy} K.~R.,  {Hester} J.~J.,   {Claussen} M.~J.,  2004, \mn@doi [\apj]
  {10.1086/421759}, \href
  {https://ui.adsabs.harvard.edu/abs/2004ApJ...610..835H} {610, 835}

\bibitem[\protect\citeauthoryear{{Henshaw}, {Caselli}, {Fontani},
  {Jim{\'e}nez-Serra}, {Tan}  \& {Hernandez}}{{Henshaw}
  et~al.}{2013}]{henshaw13}
{Henshaw} J.~D.,  {Caselli} P.,  {Fontani} F.,  {Jim{\'e}nez-Serra} I.,  {Tan}
  J.~C.,   {Hernandez} A.~K.,  2013, \mn@doi [\mnras] {10.1093/mnras/sts282},
  \href {https://ui.adsabs.harvard.edu/abs/2013MNRAS.428.3425H} {428, 3425}

\bibitem[\protect\citeauthoryear{{Hester} et~al.,}{{Hester}
  et~al.}{1996}]{hester96}
{Hester} J.~J.,  et~al., 1996, \mn@doi [\aj] {10.1086/117968}, \href
  {https://ui.adsabs.harvard.edu/abs/1996AJ....111.2349H} {111, 2349}

\bibitem[\protect\citeauthoryear{{Hunter}}{{Hunter}}{2007}]{Hunter_2007}
{Hunter} J.~D.,  2007, \mn@doi [Computing in Science and Engineering]
  {10.1109/MCSE.2007.55}, \href
  {https://ui.adsabs.harvard.edu/abs/2007CSE.....9...90H} {9, 90}

\bibitem[\protect\citeauthoryear{{Indebetouw}, {Robitaille}, {Whitney},
  {Churchwell}, {Babler}, {Meade}, {Watson}  \& {Wolfire}}{{Indebetouw}
  et~al.}{2007}]{indebetouw07}
{Indebetouw} R.,  {Robitaille} T.~P.,  {Whitney} B.~A.,  {Churchwell} E.,
  {Babler} B.,  {Meade} M.,  {Watson} C.,   {Wolfire} M.,  2007, \mn@doi [\apj]
  {10.1086/520316}, \href
  {https://ui.adsabs.harvard.edu/abs/2007ApJ...666..321I} {666, 321}

\bibitem[\protect\citeauthoryear{{Inoue} \& {Fukui}}{{Inoue} \&
  {Fukui}}{2013}]{inoue13}
{Inoue} T.,  {Fukui} Y.,  2013, \mn@doi [\apjl] {10.1088/2041-8205/774/2/L31},
  \href {https://ui.adsabs.harvard.edu/abs/2013ApJ...774L..31I} {774, L31}

\bibitem[\protect\citeauthoryear{Karim et~al.,}{Karim
  et~al.}{2023}]{karim2023sofia}
Karim R.~L.,  et~al., 2023, SOFIA FEEDBACK Survey: The Pillars of Creation in
  [C II] and Molecular Lines (\mn@eprint {arXiv} {2309.14637})

\bibitem[\protect\citeauthoryear{{Kohno} et~al.,}{{Kohno}
  et~al.}{2018}]{Kohno18}
{Kohno} M.,  et~al., 2018, \mn@doi [\pasj] {10.1093/pasj/psx137}, \href
  {https://ui.adsabs.harvard.edu/abs/2018PASJ...70S..50K} {70, S50}

\bibitem[\protect\citeauthoryear{{Kuhn}, {Hillenbrand}, {Sills}, {Feigelson}
  \& {Getman}}{{Kuhn} et~al.}{2019}]{kuhn19}
{Kuhn} M.~A.,  {Hillenbrand} L.~A.,  {Sills} A.,  {Feigelson} E.~D.,   {Getman}
  K.~V.,  2019, \mn@doi [\apj] {10.3847/1538-4357/aaef8c}, \href
  {https://ui.adsabs.harvard.edu/abs/2019ApJ...870...32K} {870, 32}

\bibitem[\protect\citeauthoryear{{Lei} et~al.,}{{Lei} et~al.}{2023}]{lei23}
{Lei} Z.,  et~al., 2023, \mn@doi [\apj] {10.3847/1538-4357/ace7b6}, \href
  {https://ui.adsabs.harvard.edu/abs/2023ApJ...954..130L} {954, 130}

\bibitem[\protect\citeauthoryear{{Linsky}, {Gagn{\'e}}, {Mytyk}, {McCaughrean}
  \& {Andersen}}{{Linsky} et~al.}{2007}]{linsky07}
{Linsky} J.~L.,  {Gagn{\'e}} M.,  {Mytyk} A.,  {McCaughrean} M.,   {Andersen}
  M.,  2007, \mn@doi [\apj] {10.1086/508763}, \href
  {https://ui.adsabs.harvard.edu/abs/2007ApJ...654..347L} {654, 347}

\bibitem[\protect\citeauthoryear{{Longair}}{{Longair}}{1992}]{longair92}
{Longair} M.~S.,  1992, {High energy astrophysics. Vol.1: Particles, photons
  and their detection}

\bibitem[\protect\citeauthoryear{{Maity}, {Dewangan}, {Sano}, {Tachihara},
  {Fukui}  \& {Bhadari}}{{Maity} et~al.}{2022}]{maity22}
{Maity} A.~K.,  {Dewangan} L.~K.,  {Sano} H.,  {Tachihara} K.,  {Fukui} Y.,
  {Bhadari} N.~K.,  2022, \mn@doi [\apj] {10.3847/1538-4357/ac7872}, \href
  {https://ui.adsabs.harvard.edu/abs/2022ApJ...934....2M} {934, 2}

\bibitem[\protect\citeauthoryear{{Maity}, {Dewangan}, {Bhadari}, {Ojha}, {Chen}
   \& {Pandey}}{{Maity} et~al.}{2023}]{maity23}
{Maity} A.~K.,  {Dewangan} L.~K.,  {Bhadari} N.~K.,  {Ojha} D.~K.,  {Chen} Z.,
   {Pandey} R.,  2023, \mn@doi [\mnras] {10.1093/mnras/stad1644}, \href
  {https://ui.adsabs.harvard.edu/abs/2023MNRAS.523.5388M} {523, 5388}

\bibitem[\protect\citeauthoryear{{Marsh}, {Whitworth}  \& {Lomax}}{{Marsh}
  et~al.}{2015}]{marsh15}
{Marsh} K.~A.,  {Whitworth} A.~P.,   {Lomax} O.,  2015, \mn@doi [\mnras]
  {10.1093/mnras/stv2248}, \href
  {https://ui.adsabs.harvard.edu/abs/2015MNRAS.454.4282M} {454, 4282}

\bibitem[\protect\citeauthoryear{{Marsh} et~al.,}{{Marsh}
  et~al.}{2017}]{marsh17}
{Marsh} K.~A.,  et~al., 2017, \mn@doi [\mnras] {10.1093/mnras/stx1723}, \href
  {https://ui.adsabs.harvard.edu/abs/2017MNRAS.471.2730M} {471, 2730}

\bibitem[\protect\citeauthoryear{{Meaburn} \& {White}}{{Meaburn} \&
  {White}}{1982}]{meaburn82}
{Meaburn} J.,  {White} N.~J.,  1982, \mn@doi [\mnras]
  {10.1093/mnras/199.1.121}, \href
  {https://ui.adsabs.harvard.edu/abs/1982MNRAS.199..121M} {199, 121}

\bibitem[\protect\citeauthoryear{{Meaburn} \& {Whitehead}}{{Meaburn} \&
  {Whitehead}}{1990}]{meaburn90}
{Meaburn} J.,  {Whitehead} M.~J.,  1990, \aap, \href
  {https://ui.adsabs.harvard.edu/abs/1990A&A...235..395M} {235, 395}

\bibitem[\protect\citeauthoryear{{Molinari} et~al.,}{{Molinari}
  et~al.}{2010a}]{Molinari10b}
{Molinari} S.,  et~al., 2010a, \mn@doi [\pasp] {10.1086/651314}, \href
  {https://ui.adsabs.harvard.edu/abs/2010PASP..122..314M} {122, 314}

\bibitem[\protect\citeauthoryear{{Molinari} et~al.,}{{Molinari}
  et~al.}{2010b}]{Molinari10a}
{Molinari} S.,  et~al., 2010b, \mn@doi [\aap] {10.1051/0004-6361/201014659},
  \href {https://ui.adsabs.harvard.edu/abs/2010A&A...518L.100M} {518, L100}

\bibitem[\protect\citeauthoryear{{Nishimura} et~al.,}{{Nishimura}
  et~al.}{2021}]{nishimura21}
{Nishimura} A.,  et~al., 2021, \mn@doi [\pasj] {10.1093/pasj/psaa083}, \href
  {https://ui.adsabs.harvard.edu/abs/2021PASJ...73S.285N} {73, S285}

\bibitem[\protect\citeauthoryear{{Oliveira}}{{Oliveira}}{2008}]{oliveira08}
{Oliveira} J.~M.,  2008, in {Reipurth} B.,  ed., , Vol.~5, Handbook of Star
  Forming Regions, Volume II.
p.~599, \mn@doi{10.48550/arXiv.0809.3735}

\bibitem[\protect\citeauthoryear{{Padovani}, {Hennebelle}, {Marcowith}  \&
  {Ferri{\`e}re}}{{Padovani} et~al.}{2015}]{padovani15}
{Padovani} M.,  {Hennebelle} P.,  {Marcowith} A.,   {Ferri{\`e}re} K.,  2015,
  \mn@doi [\aap] {10.1051/0004-6361/201526874}, \href
  {https://ui.adsabs.harvard.edu/abs/2015A&A...582L..13P} {582, L13}

\bibitem[\protect\citeauthoryear{{Padovani}, {Marcowith}, {Hennebelle}  \&
  {Ferri{\`e}re}}{{Padovani} et~al.}{2016}]{padovani16}
{Padovani} M.,  {Marcowith} A.,  {Hennebelle} P.,   {Ferri{\`e}re} K.,  2016,
  \mn@doi [\aap] {10.1051/0004-6361/201628221}, \href
  {https://ui.adsabs.harvard.edu/abs/2016A&A...590A...8P} {590, A8}

\bibitem[\protect\citeauthoryear{{Pattle} et~al.,}{{Pattle}
  et~al.}{2018}]{Pattle_2018ApJ}
{Pattle} K.,  et~al., 2018, \mn@doi [\apjl] {10.3847/2041-8213/aac771}, \href
  {https://ui.adsabs.harvard.edu/abs/2018ApJ...860L...6P} {860, L6}

\bibitem[\protect\citeauthoryear{{Pontoppidan} et~al.,}{{Pontoppidan}
  et~al.}{2022}]{pontoppidan22}
{Pontoppidan} K.~M.,  et~al., 2022, \mn@doi [\apjl] {10.3847/2041-8213/ac8a4e},
  \href {https://ui.adsabs.harvard.edu/abs/2022ApJ...936L..14P} {936, L14}

\bibitem[\protect\citeauthoryear{{Pound}}{{Pound}}{1998}]{pound98}
{Pound} M.~W.,  1998, \mn@doi [\apjl] {10.1086/311131}, \href
  {https://ui.adsabs.harvard.edu/abs/1998ApJ...493L.113P} {493, L113}

\bibitem[\protect\citeauthoryear{{Priestley} \& {Whitworth}}{{Priestley} \&
  {Whitworth}}{2021}]{Priestley21}
{Priestley} F.~D.,  {Whitworth} A.~P.,  2021, \mn@doi [\mnras]
  {10.1093/mnras/stab1777}, \href
  {https://ui.adsabs.harvard.edu/abs/2021MNRAS.506..775P} {506, 775}

\bibitem[\protect\citeauthoryear{{Reiter}, {Morse}, {Smith}, {Haworth}, {Kuhn}
  \& {Klaassen}}{{Reiter} et~al.}{2022}]{reiter22}
{Reiter} M.,  {Morse} J.~A.,  {Smith} N.,  {Haworth} T.~J.,  {Kuhn} M.~A.,
  {Klaassen} P.~D.,  2022, \mn@doi [\mnras] {10.1093/mnras/stac2820}, \href
  {https://ui.adsabs.harvard.edu/abs/2022MNRAS.517.5382R} {517, 5382}

\bibitem[\protect\citeauthoryear{{Rieke}, {Kelly}  \& {Horner}}{{Rieke}
  et~al.}{2005}]{2005SPIE.5904....1R}
{Rieke} M.~J.,  {Kelly} D.,   {Horner} S.,  2005, in {Heaney} J.~B.,
  {Burriesci} L.~G.,  eds,  Society of Photo-Optical Instrumentation Engineers
  (SPIE) Conference Series Vol. 5904, Cryogenic Optical Systems and Instruments
  XI. pp~1--8, \mn@doi{10.1117/12.615554}

\bibitem[\protect\citeauthoryear{{Rieke} et~al.,}{{Rieke}
  et~al.}{2015}]{2015PASP..127..584R}
{Rieke} G.~H.,  et~al., 2015, \mn@doi [\pasp] {10.1086/682252}, \href
  {https://ui.adsabs.harvard.edu/abs/2015PASP..127..584R} {127, 584}

\bibitem[\protect\citeauthoryear{{Rigby} et~al.,}{{Rigby}
  et~al.}{2023}]{rigby2023}
{Rigby} J.,  et~al., 2023, \mn@doi [\pasp] {10.1088/1538-3873/acb293}, \href
  {https://ui.adsabs.harvard.edu/abs/2023PASP..135d8001R} {135, 048001}

\bibitem[\protect\citeauthoryear{{Rybicki} \& {Lightman}}{{Rybicki} \&
  {Lightman}}{1979}]{rybicki79}
{Rybicki} G.~B.,  {Lightman} A.~P.,  1979, {Radiative processes in
  astrophysics}

\bibitem[\protect\citeauthoryear{{Sofue}}{{Sofue}}{2020}]{sofue20}
{Sofue} Y.,  2020, \mn@doi [\mnras] {10.1093/mnras/staa226}, \href
  {https://ui.adsabs.harvard.edu/abs/2020MNRAS.492.5966S} {492, 5966}

\bibitem[\protect\citeauthoryear{{Spitzer Science}}{{Spitzer
  Science}}{2009}]{spitzer09}
{Spitzer Science} C.,  2009, VizieR Online Data Catalog, \href
  {https://ui.adsabs.harvard.edu/abs/2009yCat.2293....0S} {p. II/293}

\bibitem[\protect\citeauthoryear{{Torii} et~al.,}{{Torii}
  et~al.}{2011}]{torii11}
{Torii} K.,  et~al., 2011, \mn@doi [\apj] {10.1088/0004-637X/738/1/46}, \href
  {https://ui.adsabs.harvard.edu/abs/2011ApJ...738...46T} {738, 46}

\bibitem[\protect\citeauthoryear{{Torii} et~al.,}{{Torii}
  et~al.}{2015}]{torii15}
{Torii} K.,  et~al., 2015, \mn@doi [\apj] {10.1088/0004-637X/806/1/7}, \href
  {https://ui.adsabs.harvard.edu/abs/2015ApJ...806....7T} {806, 7}

\bibitem[\protect\citeauthoryear{{Torii} et~al.,}{{Torii}
  et~al.}{2017}]{torii17}
{Torii} K.,  et~al., 2017, \mn@doi [\apj] {10.3847/1538-4357/835/2/142}, \href
  {https://ui.adsabs.harvard.edu/abs/2017ApJ...835..142T} {835, 142}

\bibitem[\protect\citeauthoryear{{Tremblin} et~al.,}{{Tremblin}
  et~al.}{2013}]{tremblin13}
{Tremblin} P.,  et~al., 2013, \mn@doi [\aap] {10.1051/0004-6361/201322233},
  \href {https://ui.adsabs.harvard.edu/abs/2013A&A...560A..19T} {560, A19}

\bibitem[\protect\citeauthoryear{{Umemoto} et~al.,}{{Umemoto}
  et~al.}{2017}]{umemoto17}
{Umemoto} T.,  et~al., 2017, \mn@doi [\pasj] {10.1093/pasj/psx061}, \href
  {https://ui.adsabs.harvard.edu/abs/2017PASJ...69...78U} {69, 78}

\bibitem[\protect\citeauthoryear{{White} et~al.,}{{White}
  et~al.}{1999}]{white99}
{White} G.~J.,  et~al., 1999, \aap, \href
  {https://ui.adsabs.harvard.edu/abs/1999A&A...342..233W} {342, 233}

\bibitem[\protect\citeauthoryear{{Whitworth}, {Bhattal}, {Chapman}, {Disney}
  \& {Turner}}{{Whitworth} et~al.}{1994}]{Whitworth_1994}
{Whitworth} A.~P.,  {Bhattal} A.~S.,  {Chapman} S.~J.,  {Disney} M.~J.,
  {Turner} J.~A.,  1994, \mn@doi [\mnras] {10.1093/mnras/268.1.291}, \href
  {https://ui.adsabs.harvard.edu/abs/1994MNRAS.268..291W} {268, 291}

\bibitem[\protect\citeauthoryear{{Wright} et~al.,}{{Wright}
  et~al.}{2015}]{wright2015_miri}
{Wright} G.~S.,  et~al., 2015, \mn@doi [\pasp] {10.1086/682253}, \href
  {https://ui.adsabs.harvard.edu/abs/2015PASP..127..595W} {127, 595}

\makeatother
\end{thebibliography}

\end{document}